\documentclass[aps,pra,twocolumn,groupedaddress,floatfix]{revtex4-1}

\usepackage{units}
\usepackage{graphicx}
\usepackage{amsmath}
\usepackage{float}

\begin{document}


\title{Precision measurement of the ionization energy of Cs I}


\author{Johannes Deiglmayr}
\author{Holger Herburger}
\author{Heiner Sa{\ss}mannshausen}
\author{Paul Jansen}
\author{Hansj\"urg Schmutz}
\author{Fr\'ed\'eric Merkt}

\affiliation{Laboratory of Physical Chemistry, ETH Zurich, Switzerland}
\email{jdeiglma@ethz.ch}
\email{merkt@phys.chem.ethz.ch}


\date{\today}

\begin{abstract}
We present absolute-frequency measurements for the transitions from the 6s$_{1/2}$ ground state of $^{133}$Cs to $n$p$_{1/2}$ and $n$p$_{3/2}$ Rydberg states. The transition frequencies are determined by one-photon UV spectroscopy in ultracold samples of Cs atoms using a narrowband laser system referenced to a frequency comb. From a global fit of the ionization energy $E_\mathrm{I}$ and the quantum defects of the two series we determine an improved value of $E_\mathrm{I}=\unit[h c \cdot 31\,406.467\,732\,5(14)]{cm^{-1}}$ for the ionization energy of Cs with a relative uncertainty of $5\times10^{-11}$. We also report improved values for the quantum defects of the $n$p$_{1/2}$, $n$p$_{3/2}$, $n$s$_{1/2}$, and $n$d$_{5/2}$ series.
\end{abstract}


\maketitle


Ionization energies represent important thermochemical quantities and serve as reference data to test \textit{ab-initio} quantum chemical calculations of atomic and molecular structure. Numerous methods can be employed to measure ionization energies, including photoelectron spectroscopy in various variants~\cite{muellerdethlefs98a,merkt11a}, photoionization spectroscopy~\cite{berkowitz79a,ng02a,ruscic00a}, photodetachment microscopy~\cite{blondel2005}, and Rydberg-state spectroscopy in combination with Rydberg-series extrapolation~\cite{herzberg72a}.

Over the years, the accuracy with which ionization energies can be determined experimentally has continuously improved, by more than an order of magnitude every 10 years (see Fig. 28 in Ref.~\cite{merkt11a}), approaching the accuracy of 100~MHz in the case of neutral polyatomic molecules~\cite{neuhauser97a} and singly negatively charged atoms and small molecules~\cite{blondel2005,seiler03a}, and even surpassing this accuracy in special cases such as atoms~\cite{stoicheff1979,weber1987} or molecular hydrogen~\cite{sprecher2011}. With the rapid development of methods to generate cold samples of molecules~\cite{friedrich2009,vandemeerakker2012,narevicius2012,hogan11a} and the extension of frequency combs to shorter wavelengths~\cite{kandula2010,cingoz2012}, measurements of molecular ionization energies with sub-MHz precision are becoming possible by Rydberg-state spectroscopy.

The current limit of these new tools and methods are best explored with alkali-metal atoms, because these atoms offer distinct advantages for precision measurements of ionization energies: their Rydberg states and first ionization energies can be reached by single-photon UV excitation from the ground state, \textit{i.e.} in a range where modern frequency-metrology tools can be fully exploited. Alkali-metal atoms can be easily laser cooled to sub-mK temperatures so that Doppler and transit-time broadenings become almost negligible. Finally, the closed-shell nature of the ion core implies that Rydberg series of alkali-metal atoms can be accurately treated as single ionization channels with Rydberg's formula~\cite{rydberg1890} or Ritz's formula~\cite{ritz1908}.

\begin{figure}
\includegraphics[width=0.95\linewidth]{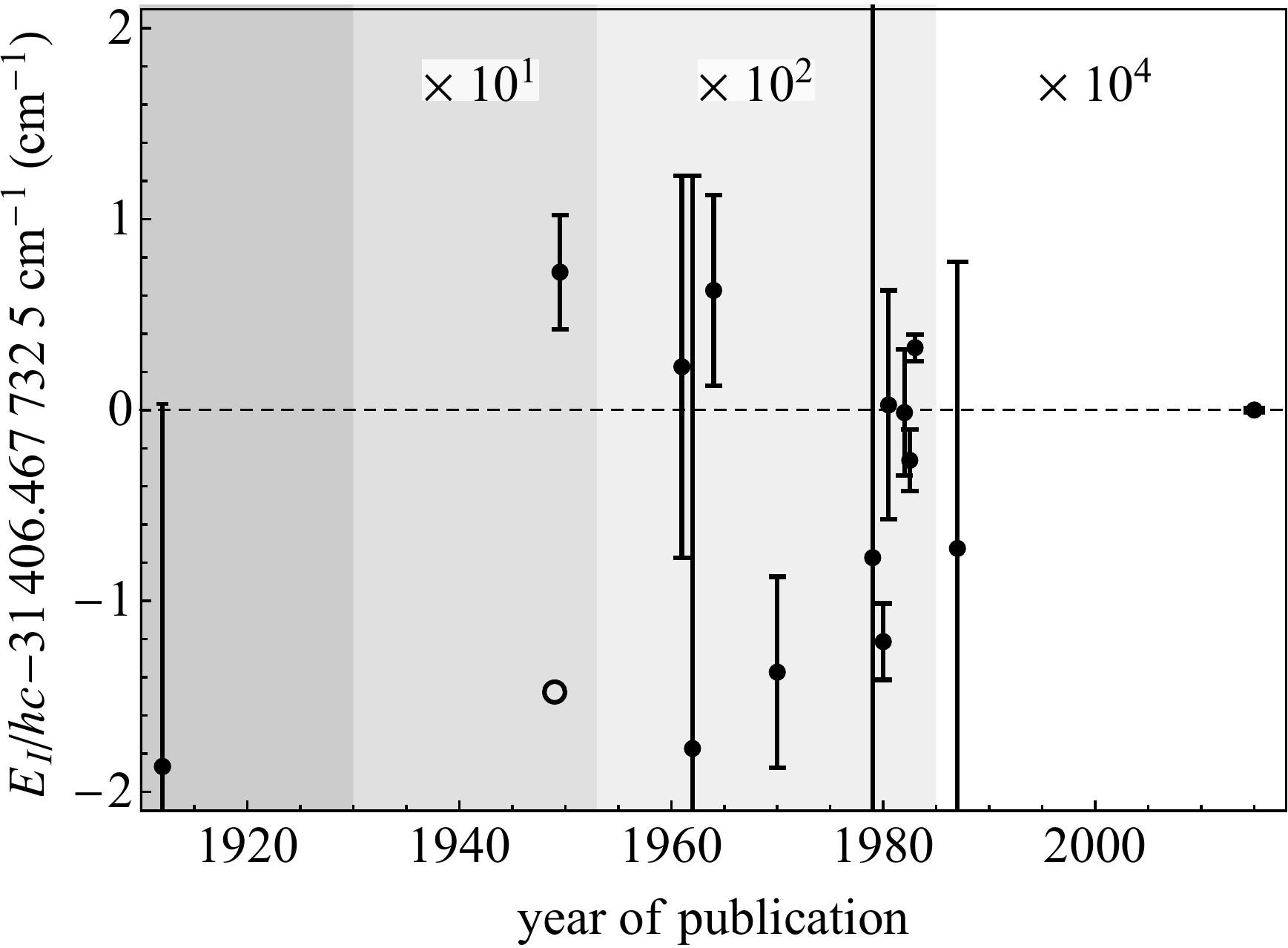}
\caption{\label{fig:historyCs} Previous measurements of the first ionization energy of $^{133}$Cs. Shown is the deviation of values reported in Ref.~\cite{sansonetti1981,lorenzen1984,osullivan1983,goy1982} from the value determined in this work (last point on the right). Error bars indicate the $1\sigma$ uncertainties given in the respective references. Open symbols indicate that no uncertainty was given. For clarity the regions marked with different shades of gray have been scaled differently as indicated at the top of each region.}
\end{figure}

Some of these advantages have been exploited in previous works: In 1979, \citet{stoicheff1979} determined the ionization energy of $^{85}$Rb with an uncertainty of 50~MHz by two-photon spectroscopy. In 2011, \citet{mack2011} determined the ionization energy of $^{87}$Rb to an accuracy of 300~kHz by referencing the excitation laser to an optical frequency comb~\cite{udem1999}. In the case of the first ionization energy of $^{133}$Cs, the progress of the experimental accuracy is summarized in Fig.~\ref{fig:historyCs}. The most accurate determination so far by \citet{weber1987}, reached an accuracy of 5~MHz in 1987.

In this article we present a new measurement of the first ionization energy of $^{133}$Cs at an accuracy of 40~kHz based on single-photon measurements of the Rydberg spectrum from the $6$s$_{1/2}$ ground state using an ultracold Cs sample and a frequency-comb-based calibration procedure.

\section{Experiment}

The experiments are performed using a sample of ultracold $^{133}$Cs atoms, released from a far-off-resonant optical dipole trap (ODT) inside an ultrahigh-vacuum chamber. The small size of typically $\sim150$~$\mu$m full width at half maximum (FWHM) of the sample in the ODT reduces the influence of electric-field gradients on the measured transition frequencies. Typical samples contain $4\cdot10^5$ atoms and have a translational temperature of 70~$\mu$K and a peak density of $6\cdot10^{10}$~cm$^{-3}$. Details of the experimental setup can be found in Ref.~\cite{sassmannshausen2013,deiglmayr2014,sassmannshausen2015}. We describe here only the aspects relevant for the measurements presented in this article. The electric field is controlled by a set of eight electrodes in the vacuum chamber, which allows for independent adjustment of the field in all three spatial directions. Stray electric fields are canceled by using a procedure based on the measurement of the quadratic Stark effect of $np_{3/2}$ Rydberg states~\cite{sassmannshausen2013}. The procedure is repeated regularly (at least daily) to ensure that residual electric fields never exceed 1~mV/cm. Similarly, stray magnetic fields are measured by radio-frequency (RF) recordings of the $F=4\leftarrow F=3$ transition in the $6s_{1/2}$ ground state of Cs and are reduced to below 2~mG by three external pairs of coils~\cite{sassmannshausen2013}.

The atoms in the ODT are first prepared in the upper hyperfine component of the 6s$_{1/2}$ ground state ($F=4$) by spatially-selective optical pumping. They are then excited in a single-photon transition to $n$p$_J$ ($J=1/2, 3/2$) Rydberg states. A Cooper minimum in the photoexcitation cross section from the 6s$_{1/2}$ ground state of Cs to the $n$p$_{1/2}$ Rydberg states just above the ionization threshold leads to a strong reduction of the absorption cross section for these states at high values of $n$~\cite{raimond1978}. We increase the power and the pulse length of the excitation laser to compensate for this effect.

For $n \ge 42$, the Rydberg atoms are detected by switching the electric potential at an additional electrode to a high value, which field ionizes the Rydberg atoms and accelerates the resulting ions toward a micro-channel-plate detector~\cite{sassmannshausen2013}. The detector signal is recorded with a digital-storage oscilloscope. The digitized trace is transferred to a computer and is analyzed by a peak-finding algorithm to determine the number of detected ions. At values of $n$ lower than 42, field-ionization is not efficient in the same configuration of applied electric potentials. We therefore introduce a delay of 100~$\mu$s between excitation and the application of the pulsed potential. During this delay, Rydberg atoms can spontaneously ionize and the resulting ions are then extracted and detected as described above. Possible ionization mechanisms include direct ionization by black-body radiation or black-body-radiation-enhanced field ionization~\cite{beterov2007}, interaction-induced Penning-ionization of pairs of Rydberg atoms~\cite{sassmannshausen2015}, and collisions between Rydberg atoms and hot atoms from the background gas in the chamber.

A ring dye laser system (Coherent 899-21), pumped by a frequency-doubled continuous-wave Nd:YVO$_4$ laser (Quantum finesse 532), and a frequency-doubling unit (Coherent MBD 200) are used to produce frequency-tunable light for the excitation into Rydberg states at wavelengths around 319~nm. The frequency of the ring dye laser is stabilized using a two-step locking scheme. First, the frequency is stabilized to an external reference cavity (Thorlabs SA200-5B). One of the cavity mirrors is mounted onto a piezoelectrical actuator which allows for tuning of the resonance frequency of the cavity. The error signal for the stabilization is derived using the Pound-Drever-Hall technique~\cite{drever1983} by modulating only a small part of the fundamental output of the ring laser with an electro-optical modulator. An analog proportional-integral (PI) controller feeds the error signal back to the electronic control box of the ring laser, which was modified according to Ref.~\cite{haubrich1996}. The closed-loop bandwidth of this lock is $\sim$8~kHz. Second, slow drifts of the length of the reference cavity are observed by measuring the laser frequency with a wave meter (HighFinesse WS-7). The measured frequency is stabilized to a chosen set point by applying a variable voltage to the piezoelectrical actuator of the reference cavity. The closed-loop bandwidth of this lock is $\sim$10~Hz. The frequency of the laser is scanned in steps of typically 150~kHz by varying the lock set point. While the absolute accuracy of the wave meter is specified as 60~MHz ($3\sigma$), the intrinsic Allan deviation of the wave meter measurements is typically around 300~kHz over periods shorter than one minute~\cite{sanguinetti2009}, which was confirmed by own measurements.

The absolute frequency of the excitation laser radiation is determined by overlapping a small part of the fundamental beam at $\sim$640~nm with light from a frequency comb (Menlo Systems FC1500-250-WG) on a fast photodiode and counting the frequency of the resulting beat note.  The absolute accuracy of the wave meter is sufficient to unambiguously determine the mode number of the beating comb tooth. All frequencies reported in this article were obtained by direct comparison with the frequency of a Rb oscillator (Stanford Research Systems FS725) having a stability of $1\times10^{-11}$ over a typical measurement time of 10~s and being disciplined by a GPS receiver (Spectrum Instruments TM-4) with a specified long-term stability of $1\times10^{-12}$. The accuracy of this clock yields directly the accuracy of the optical frequency measurements~\cite{kubina2005}. The continuous UV light is chopped into short pulses of 3 to 20~$\mu$s length using the negative first diffraction order of an acousto-optic modulator (AOM) in single-pass configuration. The RF signal driving the AOM is derived from a stable quartz oscillator at 110.000(1) MHz. Its frequency was repeatedly controlled by recording and analyzing RF-leakage signals with an antenna and a Fast-Fourier transformation on a digital oscilloscope (LeCroy WaveRunner 604Zi, specified clock accuracy at time of measurement: 4~ppm).

\section{Line-shape model and transition frequencies}\label{sec:transitionFreq}

The experimentally observed line widths are on the order of 1.2~MHz (FWHM). Possible contributions to the widths are the spontaneous decay of the Rydberg states, the Doppler broadening of the transitions, inhomogeneous broadening caused by residual electric fields, broadenings induced by Rydberg-Rydberg-interactions, the hyperfine splitting of the Rydberg states, and the bandwidth of the excitation radiation. The natural line widths of transitions to $np$ Rydberg states scale with the principal quantum number $n$ approximately as $n^{-3}$ and are less than 10~kHz for $n \ge 27$~\cite{ovsiannikov2011}. Their contributions to the experimental line widths are thus negligible. The motion of the atoms resulting from the finite temperature of the sample causes a homogeneous Doppler broadening of the transitions in the ultraviolet with an estimated FWHM of 600~kHz.
\begin{figure}
\begin{center}
\includegraphics[width=0.8\linewidth]{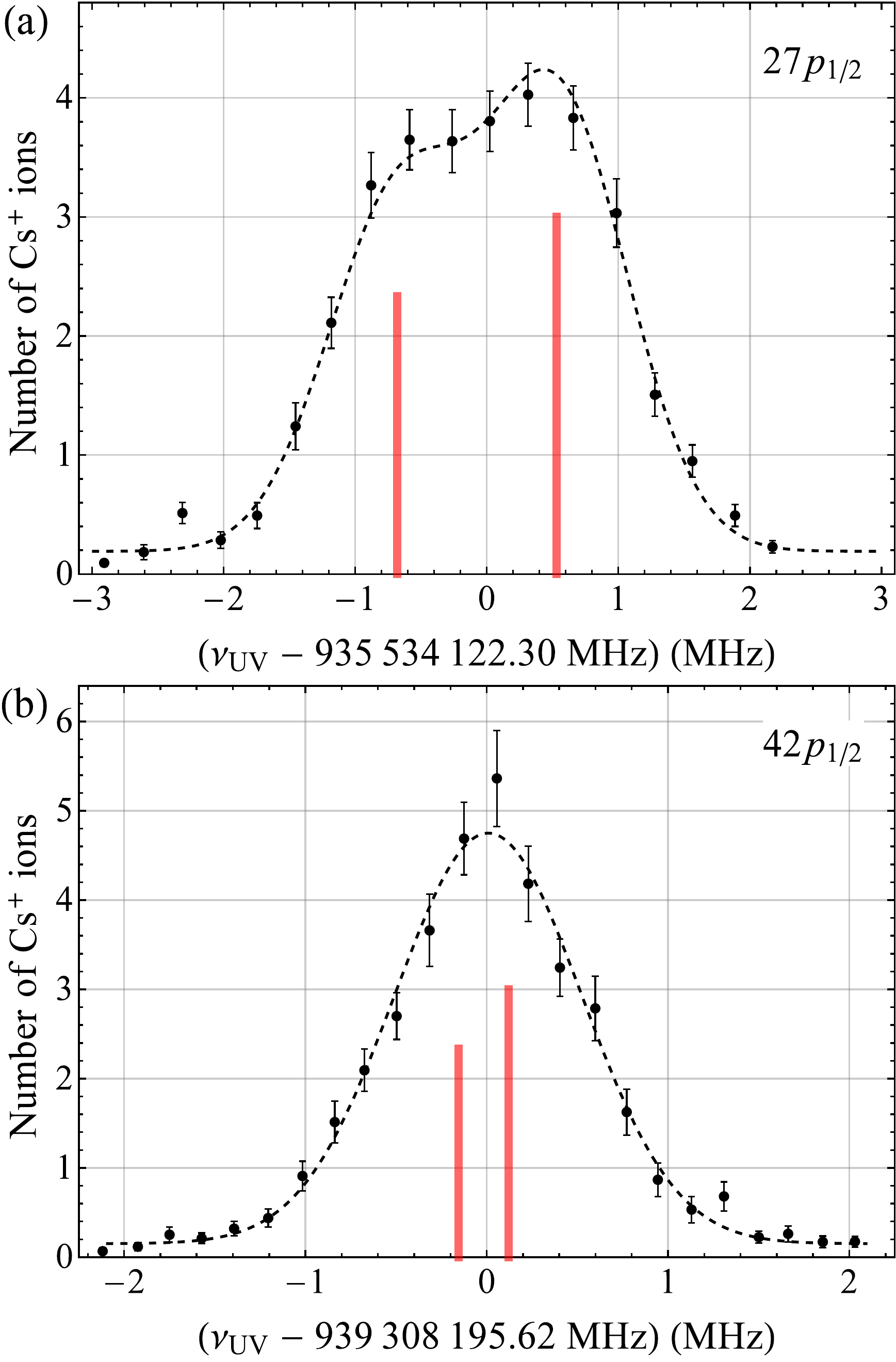}
\caption{\label{fig:resonances}(Color online) (a) Spectrum of the $27p_{1/2}\leftarrow 6$s$_{1/2}(F=4)$ transition with partially resolved hyperfine structure: (red lines) Positions and relative weights of hyperfine components,
(dashed line) fit of the line model of Eq.~\eqref{eq:line-model} to the raw data, (black points) binned raw data for visualization (error bars show the standard error of the mean). (b) Spectrum of the $42p_{1/2}\leftarrow 6$s$_{1/2}(F=4)$ transition with unresolved hyperfine structure. Legend as in (a).}
\end{center}
\end{figure}

Because the polarizability of Rydberg states with nonzero quantum defect scales as $n^7$~\cite{gallagher2005}, the inhomogeneous line broadening resulting from electric-field gradients is most severe at high $n$ values. The same holds for inhomogeneous broadenings by Van der Waals interactions between Rydberg atoms, which scale as $n^{11}$~\cite{gallagher2005} and with the Rydberg-atom density $\rho$ as $\rho^2$. After reducing the number of excited Rydberg atoms per shot to below 10 atoms, we observe an almost constant line width in the range $27\le n\le 74$. A significant contribution from these two mechanisms to the observed line width can thus be excluded.

The hyperfine interaction is a Fermi-contact-type interaction and the coupling strength scales as $n^{-3}$. In the case of Cs, the nuclear spin ($I=7/2$) and the magnetic-dipole hyperfine coupling constants are large (e.g., $A_{6s_{1/2}}=2.298\,157\,942\,5$ GHz)~\cite{Arimondo1977}, leading, at our resolution, to significant hyperfine splittings even for high Rydberg states~\cite{sassmannshausen2013}. Using experimentally determined hyperfine splittings for selected $np_{1/2}$ states from Ref.~\cite{goy1982} and for $np_{3/2}$ states from Ref.~\cite{sassmannshausen2013}, the hyperfine coupling constants $A_\textrm{hfs}$ of the investigated Rydberg states are predicted. For $27p_{1/2}$ the $F=3$ to $F=4$ interval is $1.2\pm0.1$~MHz and for $27p_{3/2}$ the $F=3$ to $F=5$ interval is $0.6\pm0.1$~MHz. The hyperfine structure must thus be taken into account explicitly at least for lower values of $n$. Weighting the different hyperfine transitions by their statistical factor of $(2F+1)$ results in the following line-shape model
\begin{align}\label{eq:line-model}
g(\nu) & \propto \sum\limits_F (2 F+1) \exp\left(-\frac{\left(\nu-\nu_F-\nu_0\right)^2}{2 \sigma^2}\right) \\ \nonumber
\nu_F & = \frac{A_\textrm{hfs}}{2}F(F+1) -\frac{A_\textrm{hfs}}{4}\left(F_<^2 + F_>\left(F_>+2\right)\right) \, ,
\end{align}
which includes all hyperfine components of a given transition. In Eq.~\eqref{eq:line-model}, $A_\textrm{hfs}$ is the magnetic-dipole coupling constant, $F_<$ ($F_>$) is the smallest (largest) allowed value of the quantum number $F$, $\nu_0$ is the center of gravity of the hyperfine structure, and the residual width $\sigma$ accounts for the finite bandwidth of the excitation laser and the Doppler width of the transition. The central frequency $\nu_0$ and the residual width $\sigma$ are the only free parameters of this model. The partially resolved hyperfine structure of the $27$p$_{1/2}\leftarrow6$s$_{1/2}(F=4)$ transition allows us to confirm the validity of the line-shape model, see Fig.~\ref{fig:resonances} (a). We therefore also employ this model to resonances with nonresolved hyperfine structures, see Fig.~\ref{fig:resonances} (b). For transitions from the 6s$_{1/2}(F=4)$ state to $n$p$_{3/2}$ states, we only consider the optically accessible hyperfine components $F=3-5$ in the line-shape model. The hyperfine shift $\nu_{6s_{1/2}(F=4)}$ of the $F=4$ hyperfine component of the 6s$_{1/2}$ ground state is taken into account by adding $\nu_{6s_{1/2}(F=4)}=4.021\,776\,4$~GHz to the observed transition frequencies.

In order to determine the precise transition frequency of a $np_{j}\leftarrow 6s_{1/2}$ resonance, we scan the laser frequency stepwise over the resonance and perform typically 150 consecutive excitation-detection cycles at each position. Without prior averaging, the model of Eq.~\eqref{eq:line-model} is fitted to the raw data using a nonlinear least-squares fitting algorithm. For several transitions, the transition frequencies are determined up to five times in separate measurements. This procedure yields the most reliable estimate of typically 60~kHz for the statistical uncertainty of the determination of transition frequencies $\nu_0$. This corresponds to approximately 1/20$^\textrm{th}$ of the typical experimental line width. The wavenumbers of all observed transitions are listed in Tab.~\ref{tab:allLines}. These values are in agreement with the respective transition frequencies reported by \citet{weber1987} within their stated uncertainties, but are more precise by about two orders of magnitude. Because \citet{weber1987} have already reviewed the agreement between their results and other previous measurements, we refer the reader to their work for a complete overview of measurements carried out before 1987.

\begin{table}
\begin{tabular}{|c|cc|cc|}
\hline
$n$ &  $\tilde{\nu}_{p 1/2}$ / cm$^{-1}$   & $\delta_\mathrm{fit}$ / kHz & $\tilde{\nu}_{p 3/2}$   / cm$^{-1}$  & $\delta_\mathrm{fit}$ / kHz \\ \hline
 27 & 31\,206.189\,769\,8 & 14   & 31\,206.744\,750\,5 & -3 \\
 30 & 31\,249.111\,102\,0 & -27  & 31\,249.497\,750\,0 & 35 \\
 33 & 31\,279.579\,380\,2 & -35  & 31\,279.859\,432\,4 & -49 \\
 36 & 31\,301.984\,579\,8 & 12   & 31\,302.193\,882\,3 & 43 \\
 39 & 31\,318.939\,791\,8 & 32   & 31\,319.100\,300\,8 & 15 \\
 42 & 31\,332.079\,304\,8 & -10  & 31\,332.205\,082\,9 & 11 \\
 45 & 31\,342.467\,758\,4 & 42   & 31\,342.568\,141\,1 & -24 \\
 47 & 31\,348.229\,448\,9 & 43   & 31\,348.316\,589\,8 & -83 \\
 50 &      --             &  --  & 31\,355.586\,932\,3 & -20 \\
 54 &      --             &  --  & 31\,363.336\,799\,2 & 22 \\
 56 & 31\,366.514\,426\,8 & -14  & 31\,366.563\,956\,6 & -18 \\
 59 & 31\,370.723\,748\,8 & 44   & 31\,370.765\,664\,1 & 49 \\
 62 & 31\,374.301\,268\,8 & -17  & 31\,374.337\,055\,9 & 59 \\
 64 & 31\,376.395\,953\,5 & -6   & 31\,376.428\,301\,8 & 9 \\
 66 & 31\,378.292\,494\,8 & -56  & 31\,378.321\,835\,7 & 51 \\
 70 & 31\,381.584\,461\,0 & -110 & 31\,381.608\,817\,4 & 70 \\
 74 & 31\,384.331\,466\,9 & -36  & 31\,384.351\,900\,1 & -17 \\\hline
\end{tabular}
\caption{\label{tab:allLines} Wavenumbers $\tilde{\nu}$ (in cm$^{-1}$) and fit residuals $\delta_\mathrm{fit}$ (in kHz, see Sect.~\ref{sec:results}) of all observed transitions to $n$p$_{1/2}$ and $n$p$_{3/2}$ Rydberg states with respect to the center of gravity of the 6s$_{1/2}$ ground state. In case a transition was measured several times we quote the mean of all determined wave numbers and fit residuals weighted by the statistical uncertainties of the fit of Eq.~\eqref{eq:line-model}. The estimated standard deviation of all line positions is $2\cdot10^{-6}$~cm$^{-1}$.}
\end{table}

\section{Systematic frequency shifts}\label{sec:systematicshifts}

\begin{figure}
\begin{center}
\includegraphics[width=0.95\linewidth]{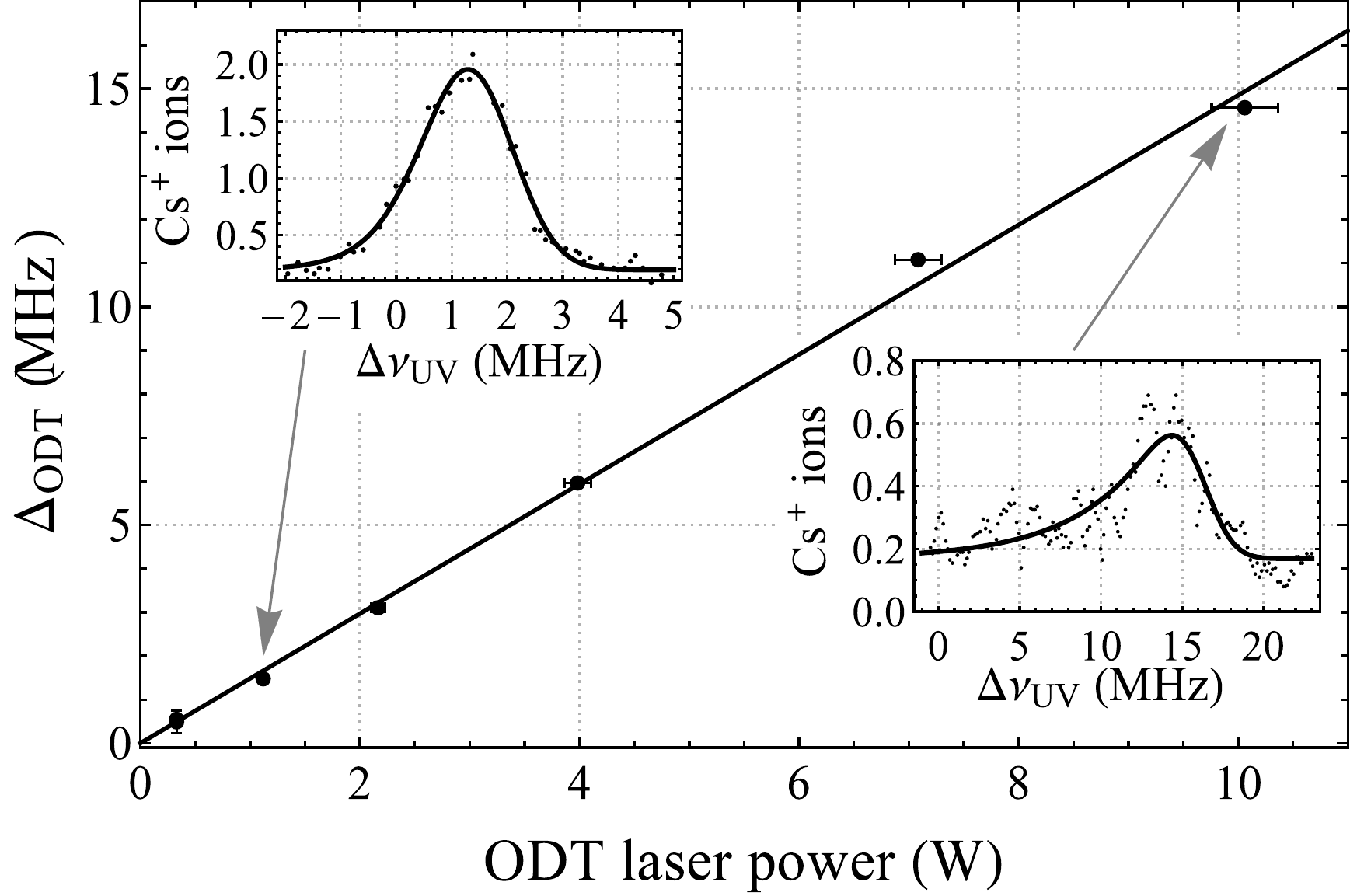}
\caption{\label{fig:ODTshift}  AC-Stark shift of the $70p_{3/2}\leftarrow6s_{1/2}$ transition of Cs measured in the ODT at different power levels of the ODT laser during excitation. Experimentally determined peak positions are displayed as full circles. The horizontal and vertical error bars indicate the estimated standard deviation of the peak position and a 3\% uncertainty in the determination of the laser power, respectively. The full line represents the fitted linear model. The upper and lower insets show experimental spectra (black points) and a fit of Eq.~\eqref{eq:skewGauss} to the spectrum (solid lines) for ODT laser powers of 1.1~W and 10.0~W, respectively. All frequencies are given relative to the extrapolated zero-power peak position.}
\end{center}
\end{figure}
All measurements are performed with samples released from an ODT. As indicated above, this approach reduces the line shifts and inhomogeneous broadenings resulting from electric-field gradients by confining the atoms to a small volume. However, for technical reasons, the power of the laser forming the ODT cannot be turned off completely but can only be reduced to a minimum value of 0.30~W. The residual optical potential leads to an AC Stark shift of the transition frequencies to higher values. We characterize and measure this shift by performing measurements at different ODT-laser powers and extrapolating the shifts to zero ODT laser power. The intensity gradient in the ODT does not only shift the resonances but also causes red-degraded line shapes which we model by an exponentially modified Gaussian line-shape function~\cite{felinger1998}
\begin{equation}\label{eq:skewGauss}
g_s(\nu) \propto \frac{1}{\tau}\exp\left(\frac{\sigma^2}{2 \tau^2}+\frac{\nu - \nu_0}{\tau}\right)\textrm{erfc}\left(\frac{\sigma}{\sqrt{2}\tau}+\frac{\nu-\nu_0}{\sqrt{2}\sigma}\right) \;.
\end{equation}
In Eq.~\eqref{eq:skewGauss}, $\nu_0$ and $\sigma$ are the central frequency and width, respectively, of the unperturbed Gaussian line profile and the parameter $\tau$ describes the line asymmetry. An exemplary set of measurements at $70p_{3/2}$ is presented in Fig.~\ref{fig:ODTshift}. The line-shape parameters are fitted in a nonlinear least-squares fit to the experimental spectral intensities and the intensity maxima are extracted from the model. For an ODT-laser power of less than $\sim 0.5$~W, the extracted position of the intensity maximum coincides with the central transition frequency obtained by the fit of Eq.~\eqref{eq:line-model} to the spectral line shape within the statistical uncertainty, justifying the use of Eq.~\eqref{eq:line-model} in the determination of the transition frequencies (Sec.~\ref{sec:transitionFreq}). By linear extrapolation of this peak position to zero ODT-laser power we determine the shift $\Delta_\textrm{ODT}$ of the transition frequency at the residual ODT-laser power of 0.30~W. We repeated this characterization measurement seven times during the data-taking period at values of $n$ in the range from 27 to 74 and did not observe any systematic variation with $n$. For the residual ODT-laser power, the ponderomotive shift of the Rydberg states and the AC-Stark shift of the ground state are estimated to be on the order of 160~kHz~\cite{liberman1983} and 320~kHz~\cite{grimm2000}, respectively. Both shifts are independent of the principal quantum number $n$, in agreement with our observations. The experimentally determined shift of a transition at the residual ODT power of 0.30~W is $\Delta_\textrm{ODT}(0.30~\textrm{W}) = 0.485(16)$~MHz, where the given uncertainty is the statistical standard deviation of the seven independent measurements. We subtract $\Delta_\textrm{ODT}$ from all measured transition frequencies.

The AC-Stark shift induced by the excitation laser was found to be negligible by varying the power of this laser while maintaining the same number of excited Rydberg atoms by either adapting the length of the excitation laser pulse or the number of ground-state atoms. The positions of the Rydberg levels are also shifted by the AC-Stark effect induced by the thermal radiation from the room-temperature environment. However, the magnitude of this effect, measured to be about 2.4~kHz at $T=300$~K~\cite{hollberg1984}, is negligible compared to the uncertainties of our measurements.

Rydberg states of atoms in dense gases experience a pressure shift resulting from collisions of the Rydberg electron with ground-state atoms located within the electron orbit~\cite{Amaldi1934,Fermi1934}. For a maximal ground-state-atom density of $10^{11}$~cm$^{-3}$ and a triplet $s$-wave scattering length of $a_T=-21.7~a_0$~\cite{bahrim2001,sassmannshausen2015}, we estimate an upper limit for the pressure shift of \unit[-13]{kHz} at high values of $n$. Reducing the ground-state-atom density by a factor of two did not lead to observable shifts of the transition frequencies at $n=74$ and we thus neglect a possible pressure shift.

Electric-field gradients and long-range Van der Waals interactions do not only lead to a line broadening, as discussed above, but also to a shift of the observed line centers. Electric-field gradients always lead to a shift of $np_j \leftarrow 6s_{1/2}$ transition frequencies to lower values, whereas the interaction-induced shift of $n$p states is positive for $n<42$ and negative for $n\ge42$~\cite{gallagher2008}. Although we did not observe systematic shifts when varying the experimental parameters (\textit{e.g.} the number of Rydberg atoms), we cannot exclude line shifts up to $\sigma_\textrm{ryd}=20$~kHz at the highest values of $n$.

\section{Ionization energy and quantum defects}\label{sec:results}

The term values of the Rydberg levels of Cs are accurately described by the extended Ritz formula~\cite{ritz1908}
\begin{align}\label{eq:ritzextended}
    \tilde{\nu}_{n \ell j} & = \frac{1}{h c}  E_\textrm{I} - \frac{R_\textrm{Cs}}{{n^*}^2} = \frac{1}{h c} E_\textrm{I} - \frac{R_\textrm{Cs}}{\left({n-\delta_{\ell j}(n)}\right)^2} \, ,
\end{align}
with
\begin{align}
    \delta_{\ell j}(n) & = \delta_{0,\ell j} + \frac{\delta_{2,\ell j}}{\left({n-\delta_{\ell j}(n)}\right)^2}+ \frac{\delta_{4,\ell j}}{\left({n-\delta_{\ell j}(n)}\right)^4} + \cdots , \label{eq:qdExpansion}
\end{align}
where $E_\textrm{I}$ is the lowest ionization energy of Cs, $R_\textrm{Cs}$ is the reduced Rydberg constant of Cs, and $\delta_{\ell j}(n)$ are the energy-dependent quantum defects of the respective series. Using the currently recommended values of fundamental constants~\cite{Nist2014} and the Cs mass~\cite{CIAAW2015} we calculate $R_\textrm{Cs}=\unit[109\,736.862\,733\,9(6)]{cm^{-1}}$. We performed a global fit of Eq.~\eqref{eq:ritzextended} to all observed transitions (weighted by their statistical uncertainties) by truncating the expansion of the energy-dependent quantum defects after the linear term:
\begin{align}\label{eq:QDglobalFit}
    \delta_{\ell j}(n) & = \delta_{0,\ell j} + \frac{\delta_{2,\ell j}}{\left({n-\delta_{0,\ell j}}\right)^2} \;.
\end{align}
Note the replacement of $\delta_{\ell j}(n)$ by $\delta_{0, \ell j}$ in the denominator of the $\delta_2$ term. As \citet{drake_1991} discuss, this modified expression allows for a simultaneous fit of the quantum defects and the ionization energy at the cost of a loss of physical meaning for the expansion coefficients $\delta_{k,\ell j}$. However we verified that an iterative fit of the quantum defects using Eq.~\eqref{eq:qdExpansion} (restricted in the expansion to $\delta_0$ and $\delta_2$) to the data leads to results identical to the ones obtained by the global fit using Eq.~\eqref{eq:QDglobalFit} within the experimental uncertainty. We also verified that the inclusion of higher-order terms in Eq.~\eqref{eq:qdExpansion} (\textit{i.e.}, $\delta_{4,\ell j}$, $\delta_{6,\ell j}$, \dots) does not reduce the residuals of the global fit.

\begin{table}
\begin{center}
\begin{tabular}{|c|c|c|}\hline
            &   $n$p$_{1/2}$  & $n$p$_{3/2}$ \\ \hline
$\delta_0$  &  3.591\,587\,1(3)   & 3.559\,067\,6(3)  \\
$\delta_2$  &  0.362\,73(16)      & 0.374\,69(14)   \\ \hline
\end{tabular}
\end{center}
\caption{\label{tab:qdpstates}Quantum-defect expansion coefficients for the $n$p series of Cs determined from a global fit to the observed transitions. The uncertainties are the statistical standard deviations extracted from the global fit.}
\end{table}

\begin{figure}
\includegraphics[width=\linewidth]{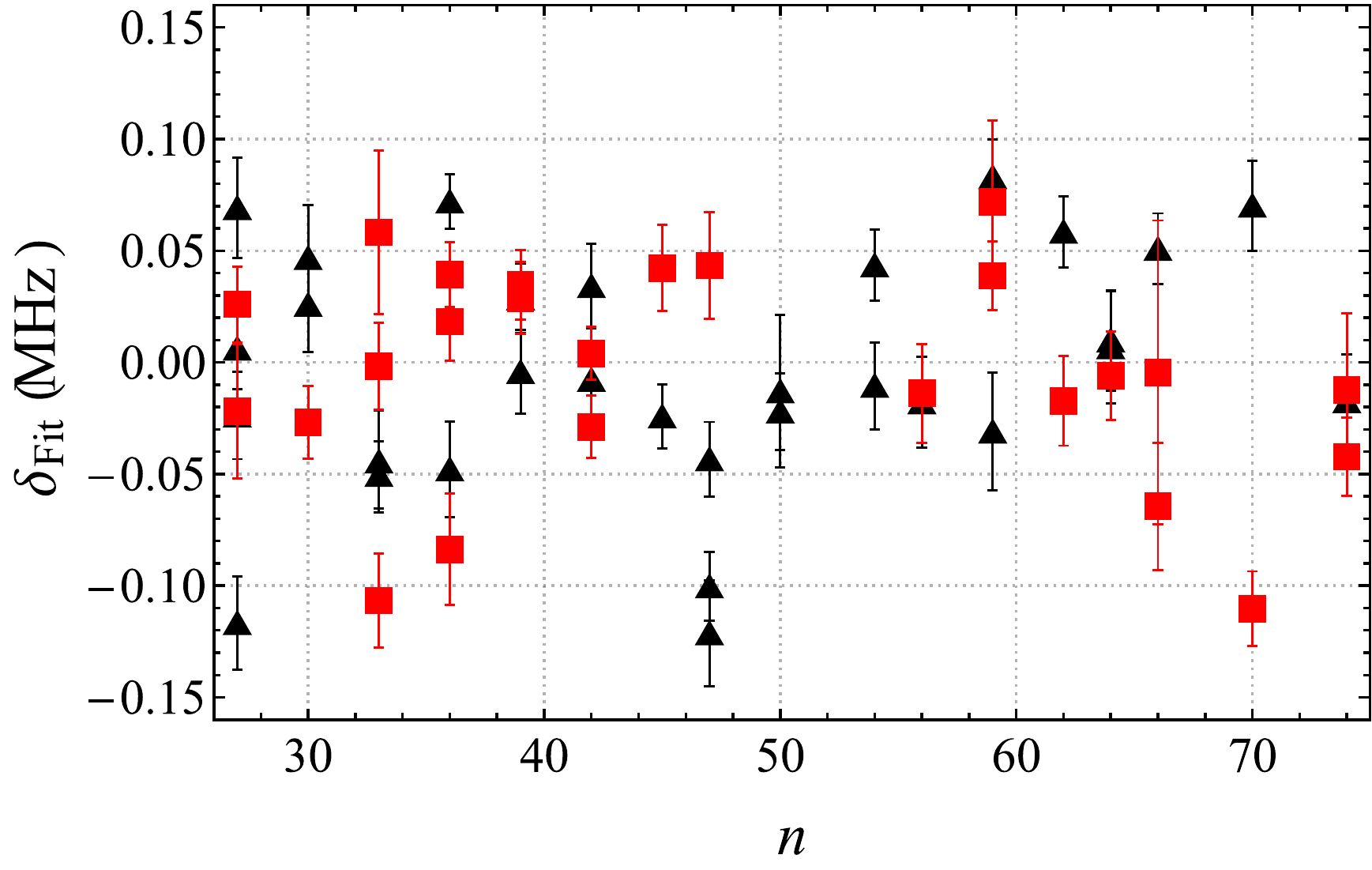}
\caption{\label{fig:residuals} (Color online) Fit residuals of the global fit based on Eq.~\eqref{eq:ritzextended} and \eqref{eq:QDglobalFit} to the observed frequencies of the $n$p$_{3/2}\leftarrow 6$s$_{1/2}$ (black triangles) and $n$p$_{1/2}\leftarrow 6$s$_{1/2}$ (red squares) transitions. The error bars give the estimated standard deviations of the center frequencies resulting from the fit of Eq.~\eqref{eq:line-model} to the raw data (see Sec.~\ref{sec:transitionFreq}).}
\end{figure}

The first ionization energy of Cs resulting from the fit is $E_\mathrm{I}=h c \cdot \unit[31\,406.467\,732\,5(14)]{cm^{-1}}$, where the quoted uncertainty is found by adding $\sigma_\textrm{ryd}$ and the statistical uncertainties of $E_\textrm{I}$ and $\Delta_\textrm{ODT}$ in quadrature. This result is in agreement with the ionization energy reported by \citet{weber1987} ($\unit[h c \cdot 31\,406.467\,66(15)]{cm^{-1}}$), however its uncertainty is reduced by two orders of magnitude. The parameters of the expansion of the quantum defect (Eq.~\eqref{eq:QDglobalFit}) are given in Tab.~\ref{tab:qdpstates}. A direct comparison with previously reported values is difficult because of the different orders in the series expansion of the quantum defects. However, we observe that the transition frequencies predicted by combining our values for the ionization energy with the quantum defects reported in Ref.~\cite{weber1987} for the $n$p$_{1/2}$ series deviate from our experimental observations by up to 200~kHz for the lowest $n$ $(\sim 30)$, which is significantly more than the residuals obtained with the quantum defects of Tab.~\ref{tab:qdpstates} (see Fig.~\ref{fig:residuals}).

All residuals $\delta_\mathrm{fit}$ of the global fit (see Fig.~\ref{fig:residuals}) are below 10\% of the experimental line width. The standard deviations of the transition frequencies, extracted from the fits of Eq.~\eqref{eq:line-model} to the raw data, seem to underestimate the true uncertainties by roughly a factor of two. They are nevertheless a good measure of the quality of the fit, justifying their use as weights in the global fit. The standardized fit residuals are almost symmetrically distributed around zero, which indicates that the systematic shifts discussed in Sec.~\ref{sec:systematicshifts} are small. The estimated statistical uncertainties of the parameters obtained from the global fit are dominated by correlations between the values of the parameters. If we use the quantum defects obtained from the global fit to calculate the ionization energy separately from every observed transition using Eq.~\eqref{eq:ritzextended}, we obtain an estimate for the error of the average value, which is about $\unit[2.5\cdot10^{-7}]{cm^{-1}}$, \textit{i.e.} four times smaller than the uncertainty of $E_\mathrm{I}$ resulting from the global fit.

In a previous article~\cite{sassmannshausen2013} we have reported frequency measurements of interseries transitions (\textit{i.e.} $\mathrm{p}_{3/2}\rightarrow \mathrm{s}_{1/2}$ and $\mathrm{p}_{3/2}\rightarrow \mathrm{d}_{3/2,5/2}$) using millimeter-wave radiation. In Tab.~\ref{tab:mmwave} we summarize the values of selected intervals for which the transition frequency relative to the center of gravity of the hyperfine-split resonances could be determined with an absolute accuracy of 10~kHz. By combining these intervals, the observed transition frequencies to the $n$p$_{3/2}$ states from Tab.~\ref{tab:allLines}, and the ionization energy determined above, we obtain the binding energies $t^b_{n,\ell j} = \tilde{\nu}_{n \ell j} - \frac{E_\textrm{I}}{h c}$ of several $n$s$_{1/2}$ and $n$d$_{5/2}$ Rydberg states listed in Tab.~\ref{tab:mmwave}. Binding energies are less sensitive to absolute calibration errors than transition frequencies and are best suited to combine measurements from different sources. We extract the binding energies for the transitions to the $n$s$_{1/2}$ and $n$d$_{5/2}$ series reported in Ref.~\cite{weber1987} by subtracting the ionization energy of the respective series from the reported transition energies. Using these binding energies and the binding energies of Tab.~\ref{tab:mmwave} with their respective uncertainties ($\sigma=2\cdot10^{-4}$~cm$^{-1}$ for data from Ref.~\cite{weber1987} and $\sigma=2\cdot10^{-6}$~cm$^{-1}$ for our data), we determine the energy dependence of the quantum defects (Eq.~\eqref{eq:qdExpansion}) in a nonlinear least-squares fit. To reach convergence of the fit residuals, it was necessary to include higher-order terms up to $\delta_{8}$ in the expansion. The values of the fitted parameters are listed in Tab.~\ref{tab:qdsdstates}. These parameters simultaneously reproduce our data (Tab.~\ref{tab:mmwave}) within the experimental uncertainty and the transition frequencies of Ref.~\cite{weber1987} with a similar sum of squared errors as the parameters reported in Ref.~\cite{weber1987}.

\begin{table}
\begin{center}
\begin{tabular}{|c|c|c|}\hline
Transition $f\leftarrow i$  &   Interval / MHz & $t^b_{{n_f},{\ell_f} {j_f}}$ / cm$^{-1}$\\ \hline
$49$s$_{1/2} \leftarrow 45$p$_{3/2}$ & 287\,476.992(10) &  -54.310\,390\,4\\ \hline
$68$s$_{1/2} \leftarrow 59$p$_{3/2}$ & 265\,898.688(10) &  -26.832\,644\,6\\ \hline
$81$s$_{1/2} \leftarrow 67$p$_{3/2}$ & 261\,818.142(10) &  -18.532\,264\,3\\ \hline
$90$s$_{1/2} \leftarrow 72$p$_{3/2}$ & 257\,008.756(10) &  -14.854\,382\,7\\ \hline
$66$d$_{5/2} \leftarrow 59$p$_{3/2}$ & 255\,306.920(10) &  -27.185\,947\,9\\ \hline
\end{tabular}
\end{center}
\caption{\label{tab:mmwave}Interseries intervals from Ref.~\cite{sassmannshausen2013} determined by high-resolution millimeter-wave spectroscopy,
and binding energies of the final Rydberg state obtained as described in the text.
}
\end{table}

\begin{table}
\begin{center}
\begin{tabular}{|c|c|c|}\hline
            &   $n$s$_{1/2}$    & $n$d$_{5/2}$ \\ \hline
$\delta_0$  &  4.049\,353\,2(4) & 2.466\,314\,4(6)  \\
$\delta_2$  &  0.239\,1(5)      & 0.013\,81(15)   \\
$\delta_4$  &  0.06(10)       & -0.392(12)   \\
$\delta_6$  &  11(7)            & -1.9(3)   \\
$\delta_8$  &  -209(150)        &    \\ \hline
\end{tabular}
\end{center}
\caption{\label{tab:qdsdstates} Quantum defects of the $n$s$_{1/2}$ and $n$d$_{5/2}$ series obtained by combining data from Ref.~\cite{weber1987} (for $11\le n\le 31$ and $9\le n\le 36$, respectively) and our measurements (see text for details). The quoted uncertainties are the estimated standard deviations from the fit.}
\end{table}

\section{Conclusions and outlook}

We have presented an experiment with which the lowest ionization energy of $^{133}$Cs was determined to be $E_\mathrm{I}=\unit[h c \cdot 31\,406.467\,732\,5(14)]{cm^{-1}}$ with a relative uncertainty of $5\cdot10^{-11}$. The uncertainty is two orders of magnitude smaller than the best previous result~\cite{weber1987} and is limited by experimental sources of error and to a lesser extent by correlations of the parameters in the quantum-defect model of the Rydberg-atom term values and the ionization energy. The experimental uncertainties, especially those resulting from the laser bandwidth and from the determination of the residual AC Stark shifts of the atoms in the ODT, might be reduced by a factor of at least three by using a UV laser with a narrower bandwidth than the laser available for the present investigation, and by performing the experiment with a small sample of atoms in a completely field-free environment. The correlations of the parameters in the global fit might be reduced by increasing the number of fitted Rydberg series to include, \textit{e.g.}, $n$s$_{1/2}$ and $n$d$_{3/2,5/2}$, for which unfortunately no data of comparable precision is available. Our analysis also resulted in improved energy-dependent quantum defects for the $n$s$_{1/2}$, $n$p$_{1/2}$, $n$p$_{3/2}$, and $n$d$_{5/2}$ series of Cs.

With the present results, we demonstrate that ionization energies can be determined with a precision of $1.4\cdot10^{-6}$~cm$^{-1}$ (42~kHz) by combining (ultra)cold atoms, frequency-comb-based calibration and Rydberg-series extrapolation. Given the improvement margins of the experiment, a precision on the order of 10~kHz seems possible in the near future.

Cs presents distinct advantages for a precise measurement of ionization  energies. It can be easily laser cooled to sub-mK temperatures; its first ionization threshold is reachable from the ground state with a single UV photon; its large mass reduces the Doppler broadening, and the closed-shell nature of the Cs$^+$ ion core facilitates the Rydberg-series extrapolation. It was therefore an ideal system to test the precision limits of the experiment. Unfortunately, Cs possesses too many electrons for being an attractive system for accurate \textit{ab-initio} quantum-chemical calculations in the near future.

On the basis of the results presented in this article we anticipate that new methods of generating cold samples of few-electron molecules~\cite{seiler2011b,motsch2014} and advances towards extending frequency combs to the far-UV range of the electromagnetic spectrum~\cite{kandula2010,cingoz2012} will soon permit measurements of few-electron molecules with a similar accuracy. The most accurate determination of the ionization energy in a molecular system, H$_2$, has an uncertainty of 12~MHz~\cite{liu2009} and has stimulated advances in the \textit{ab-initio} calculations of molecular energies including adiabatic, nonadiabatic, relativistic and quantum-electrodynamics corrections~\cite{piszczatowski2009,pachucki2010,pachucki2014}. Such calculations can now reach an accuracy comparable to that of the experiments. The uncertainty in the electron-proton mass ratio (currently $5.446\,170\,213\,52(52)\cdot10^{-4}$, \textit{i.e.}, a fractional uncertainty of $9.5\cdot10^{-11}$~\cite{Nist2014}) imposes a fundamental limit of a few kHz to the accuracy of theoretical determinations of the ionization energy of H$_2$~\cite{sprecher2011}. Precision spectroscopy of Rydberg states of H$_2$ in combination with Rydberg-series extrapolation has the potential to be even more accurate.

\begin{acknowledgments}
This work is supported financially by the Swiss National Science Foundation under Project Nr.~200020-159848, the NCCR QSIT, and the EU Initial Training Network COHERENCE under grant FP7-PEOPLE-2010-ITN-265031. We acknowledge the European Union H2020 FET Proactive project RySQ (grant N. 640378). P.J. acknowledges support through an ETH fellowship.
\end{acknowledgments}


\begin{thebibliography}{55}%
\makeatletter
\providecommand \@ifxundefined [1]{%
 \@ifx{#1\undefined}
}%
\providecommand \@ifnum [1]{%
 \ifnum #1\expandafter \@firstoftwo
 \else \expandafter \@secondoftwo
 \fi
}%
\providecommand \@ifx [1]{%
 \ifx #1\expandafter \@firstoftwo
 \else \expandafter \@secondoftwo
 \fi
}%
\providecommand \natexlab [1]{#1}%
\providecommand \enquote  [1]{``#1''}%
\providecommand \bibnamefont  [1]{#1}%
\providecommand \bibfnamefont [1]{#1}%
\providecommand \citenamefont [1]{#1}%
\providecommand \href@noop [0]{\@secondoftwo}%
\providecommand \href [0]{\begingroup \@sanitize@url \@href}%
\providecommand \@href[1]{\@@startlink{#1}\@@href}%
\providecommand \@@href[1]{\endgroup#1\@@endlink}%
\providecommand \@sanitize@url [0]{\catcode `\\12\catcode `\$12\catcode
  `\&12\catcode `\#12\catcode `\^12\catcode `\_12\catcode `\%12\relax}%
\providecommand \@@startlink[1]{}%
\providecommand \@@endlink[0]{}%
\providecommand \url  [0]{\begingroup\@sanitize@url \@url }%
\providecommand \@url [1]{\endgroup\@href {#1}{\urlprefix }}%
\providecommand \urlprefix  [0]{URL }%
\providecommand \Eprint [0]{\href }%
\providecommand \doibase [0]{http://dx.doi.org/}%
\providecommand \selectlanguage [0]{\@gobble}%
\providecommand \bibinfo  [0]{\@secondoftwo}%
\providecommand \bibfield  [0]{\@secondoftwo}%
\providecommand \translation [1]{[#1]}%
\providecommand \BibitemOpen [0]{}%
\providecommand \bibitemStop [0]{}%
\providecommand \bibitemNoStop [0]{.\EOS\space}%
\providecommand \EOS [0]{\spacefactor3000\relax}%
\providecommand \BibitemShut  [1]{\csname bibitem#1\endcsname}%
\let\auto@bib@innerbib\@empty
\bibitem [{\citenamefont {M{\"u}ller-Dethlefs}\ and\ \citenamefont
  {Schlag}(1998)}]{muellerdethlefs98a}%
  \BibitemOpen
  \bibfield  {author} {\bibinfo {author} {\bibfnamefont {K.}~\bibnamefont
  {M{\"u}ller-Dethlefs}}\ and\ \bibinfo {author} {\bibfnamefont {E.~W.}\
  \bibnamefont {Schlag}},\ }\href {\doibase
  10.1002/(SICI)1521-3773(19980605)37:10<1346::AID-ANIE1346>3.0.CO;2-H}
  {\bibfield  {journal} {\bibinfo  {journal} {Angew. Chem. (int. ed. engl.)}\
  }\textbf {\bibinfo {volume} {37}},\ \bibinfo {pages} {1346} (\bibinfo {year}
  {1998})}\BibitemShut {NoStop}%
\bibitem [{\citenamefont {Merkt}\ \emph {et~al.}(2011)\citenamefont {Merkt},
  \citenamefont {Willitsch},\ and\ \citenamefont {Hollenstein}}]{merkt11a}%
  \BibitemOpen
  \bibfield  {author} {\bibinfo {author} {\bibfnamefont {F.}~\bibnamefont
  {Merkt}}, \bibinfo {author} {\bibfnamefont {S.}~\bibnamefont {Willitsch}}, \
  and\ \bibinfo {author} {\bibfnamefont {U.}~\bibnamefont {Hollenstein}},\ }in\
  \href {\doibase 10.1002/9780470749593.hrs071} {\emph {\bibinfo {booktitle}
  {{H}andbook of {H}igh-{R}esolution {S}pectroscopy}}},\ Vol.~\bibinfo {volume}
  {3},\ \bibinfo {editor} {edited by\ \bibinfo {editor} {\bibfnamefont
  {M.}~\bibnamefont {Quack}}\ and\ \bibinfo {editor} {\bibfnamefont
  {F.}~\bibnamefont {Merkt}}}\ (\bibinfo  {publisher} {John Wiley \& Sons},\
  \bibinfo {address} {Chichester},\ \bibinfo {year} {2011})\ pp.\ \bibinfo
  {pages} {1617--1654}\BibitemShut {NoStop}%
\bibitem [{\citenamefont {Berkowitz}(1979)}]{berkowitz79a}%
  \BibitemOpen
  \bibfield  {author} {\bibinfo {author} {\bibfnamefont {J.}~\bibnamefont
  {Berkowitz}},\ }\href@noop {} {\emph {\bibinfo {title} {Photoabsorption,
  photoionization and photoelectron spectroscopy}}}\ (\bibinfo  {publisher}
  {Academic Press},\ \bibinfo {address} {New York},\ \bibinfo {year}
  {1979})\BibitemShut {NoStop}%
\bibitem [{\citenamefont {Ng}(2002)}]{ng02a}%
  \BibitemOpen
  \bibfield  {author} {\bibinfo {author} {\bibfnamefont {C.-Y.}\ \bibnamefont
  {Ng}},\ }\href {\doibase 10.1146/annurev.physchem.53.082001.144416}
  {\bibfield  {journal} {\bibinfo  {journal} {Ann. Rev. Phys. Chem.}\ }\textbf
  {\bibinfo {volume} {53}},\ \bibinfo {pages} {101} (\bibinfo {year}
  {2002})}\BibitemShut {NoStop}%
\bibitem [{\citenamefont {Ruscic}(2000)}]{ruscic00a}%
  \BibitemOpen
  \bibfield  {author} {\bibinfo {author} {\bibfnamefont {B.}~\bibnamefont
  {Ruscic}},\ }in\ \href@noop {} {\emph {\bibinfo {booktitle} {Research
  Advances in Physical Chemistry}}},\ Vol.~\bibinfo {volume} {1},\ \bibinfo
  {editor} {edited by\ \bibinfo {editor} {\bibfnamefont {R.~M.}\ \bibnamefont
  {Mohan}}}\ (\bibinfo  {publisher} {Global},\ \bibinfo {address} {Trivandrum,
  India},\ \bibinfo {year} {2000})\ pp.\ \bibinfo {pages} {39--75}\BibitemShut
  {NoStop}%
\bibitem [{\citenamefont {Blondel}\ \emph {et~al.}(2005)\citenamefont
  {Blondel}, \citenamefont {{Chaibi}}, \citenamefont {{Delsart}}, \citenamefont
  {{Drag}}, \citenamefont {{Goldfarb}},\ and\ \citenamefont
  {{Kr}{\"o}ger}}]{blondel2005}%
  \BibitemOpen
  \bibfield  {author} {\bibinfo {author} {\bibfnamefont {C.}~\bibnamefont
  {Blondel}}, \bibinfo {author} {\bibfnamefont {W.}~\bibnamefont {{Chaibi}}},
  \bibinfo {author} {\bibfnamefont {C.}~\bibnamefont {{Delsart}}}, \bibinfo
  {author} {\bibfnamefont {C.}~\bibnamefont {{Drag}}}, \bibinfo {author}
  {\bibfnamefont {F.}~\bibnamefont {{Goldfarb}}}, \ and\ \bibinfo {author}
  {\bibfnamefont {S.}~\bibnamefont {{Kr}{\"o}ger}},\ }\href@noop {} {\bibfield
  {journal} {\bibinfo  {journal} {{Eur.} {Phys.} {J.} {D}}\ }\textbf {\bibinfo
  {volume} {33}},\ \bibinfo {pages} {335} (\bibinfo {year} {2005})}\BibitemShut
  {NoStop}%
\bibitem [{\citenamefont {Herzberg}\ and\ \citenamefont
  {Jungen}(1972)}]{herzberg72a}%
  \BibitemOpen
  \bibfield  {author} {\bibinfo {author} {\bibfnamefont {G.}~\bibnamefont
  {Herzberg}}\ and\ \bibinfo {author} {\bibfnamefont
  {{\mbox{Ch}}.}~\bibnamefont {Jungen}},\ }\href {\doibase
  10.1016/0022-2852(72)90064-1} {\bibfield  {journal} {\bibinfo  {journal} {J.
  Mol. Spec.}\ }\textbf {\bibinfo {volume} {41}},\ \bibinfo {pages} {425}
  (\bibinfo {year} {1972})}\BibitemShut {NoStop}%
\bibitem [{\citenamefont {Neuhauser}\ \emph {et~al.}(1997)\citenamefont
  {Neuhauser}, \citenamefont {Siglow},\ and\ \citenamefont
  {Neusser}}]{neuhauser97a}%
  \BibitemOpen
  \bibfield  {author} {\bibinfo {author} {\bibfnamefont {R.~G.}\ \bibnamefont
  {Neuhauser}}, \bibinfo {author} {\bibfnamefont {K.}~\bibnamefont {Siglow}}, \
  and\ \bibinfo {author} {\bibfnamefont {H.~J.}\ \bibnamefont {Neusser}},\
  }\href {\doibase 10.1063/1.473170} {\bibfield  {journal} {\bibinfo  {journal}
  {J. Chem. Phys.}\ }\textbf {\bibinfo {volume} {106}},\ \bibinfo {pages} {896}
  (\bibinfo {year} {1997})}\BibitemShut {NoStop}%
\bibitem [{\citenamefont {Seiler}\ \emph {et~al.}(2003)\citenamefont {Seiler},
  \citenamefont {Hollenstein}, \citenamefont {Softley},\ and\ \citenamefont
  {Merkt}}]{seiler03a}%
  \BibitemOpen
  \bibfield  {author} {\bibinfo {author} {\bibfnamefont {R.}~\bibnamefont
  {Seiler}}, \bibinfo {author} {\bibfnamefont {U.}~\bibnamefont {Hollenstein}},
  \bibinfo {author} {\bibfnamefont {T.~P.}\ \bibnamefont {Softley}}, \ and\
  \bibinfo {author} {\bibfnamefont {F.}~\bibnamefont {Merkt}},\ }\href
  {\doibase 10.1063/1.1571528} {\bibfield  {journal} {\bibinfo  {journal} {J.
  Chem. Phys.}\ }\textbf {\bibinfo {volume} {118}},\ \bibinfo {pages} {10024}
  (\bibinfo {year} {2003})}\BibitemShut {NoStop}%
\bibitem [{\citenamefont {Stoicheff}\ and\ \citenamefont
  {{Weinberger}}(1979)}]{stoicheff1979}%
  \BibitemOpen
  \bibfield  {author} {\bibinfo {author} {\bibfnamefont {B.~P.}\ \bibnamefont
  {Stoicheff}}\ and\ \bibinfo {author} {\bibfnamefont {E.}~\bibnamefont
  {{Weinberger}}},\ }\href@noop {} {\bibfield  {journal} {\bibinfo  {journal}
  {Can. {J.} {Phys.}}\ }\textbf {\bibinfo {volume} {57}},\ \bibinfo {pages}
  {2143} (\bibinfo {year} {1979})}\BibitemShut {NoStop}%
\bibitem [{\citenamefont {Weber}\ and\ \citenamefont
  {Sansonetti}(1987)}]{weber1987}%
  \BibitemOpen
  \bibfield  {author} {\bibinfo {author} {\bibfnamefont {K.~H.}\ \bibnamefont
  {Weber}}\ and\ \bibinfo {author} {\bibfnamefont {C.~J.}\ \bibnamefont
  {Sansonetti}},\ }\href@noop {} {\bibfield  {journal} {\bibinfo  {journal}
  {Phys. Rev. A}\ }\textbf {\bibinfo {volume} {35}},\ \bibinfo {pages} {4650}
  (\bibinfo {year} {1987})}\BibitemShut {NoStop}%
\bibitem [{\citenamefont {Sprecher}\ \emph {et~al.}(2011)\citenamefont
  {Sprecher}, \citenamefont {{Jungen}}, \citenamefont {{Ubachs}},\ and\
  \citenamefont {{Merkt}}}]{sprecher2011}%
  \BibitemOpen
  \bibfield  {author} {\bibinfo {author} {\bibfnamefont {D.}~\bibnamefont
  {Sprecher}}, \bibinfo {author} {\bibfnamefont {C.}~\bibnamefont {{Jungen}}},
  \bibinfo {author} {\bibfnamefont {W.}~\bibnamefont {{Ubachs}}}, \ and\
  \bibinfo {author} {\bibfnamefont {F.}~\bibnamefont {{Merkt}}},\ }\href@noop
  {} {\bibfield  {journal} {\bibinfo  {journal} {Faraday Discuss.}\ }\textbf
  {\bibinfo {volume} {150}},\ \bibinfo {pages} {51} (\bibinfo {year}
  {2011})}\BibitemShut {NoStop}%
\bibitem [{\citenamefont {Friedrich}\ and\ \citenamefont
  {{Doyle}}(2009)}]{friedrich2009}%
  \BibitemOpen
  \bibfield  {author} {\bibinfo {author} {\bibfnamefont {B.}~\bibnamefont
  {Friedrich}}\ and\ \bibinfo {author} {\bibfnamefont {J.~M.}\ \bibnamefont
  {{Doyle}}},\ }\href@noop {} {\bibfield  {journal} {\bibinfo  {journal}
  {{Chem. Phys. Chem.}}\ }\textbf {\bibinfo {volume} {10}},\ \bibinfo {pages}
  {604} (\bibinfo {year} {2009})}\BibitemShut {NoStop}%
\bibitem [{\citenamefont {{van de Meerakker}}\ \emph
  {et~al.}(2012)\citenamefont {{van de Meerakker}}, \citenamefont {{Bethlem}},
  \citenamefont {{Vanhaecke}},\ and\ \citenamefont
  {{Meijer}}}]{vandemeerakker2012}%
  \BibitemOpen
  \bibfield  {author} {\bibinfo {author} {\bibfnamefont {S.~Y.~T.}\
  \bibnamefont {{van de Meerakker}}}, \bibinfo {author} {\bibfnamefont {H.~L.}\
  \bibnamefont {{Bethlem}}}, \bibinfo {author} {\bibfnamefont {N.}~\bibnamefont
  {{Vanhaecke}}}, \ and\ \bibinfo {author} {\bibfnamefont {G.}~\bibnamefont
  {{Meijer}}},\ }\href@noop {} {\bibfield  {journal} {\bibinfo  {journal}
  {Chem. {Rev.}}\ }\textbf {\bibinfo {volume} {112}},\ \bibinfo {pages} {4828}
  (\bibinfo {year} {2012})}\BibitemShut {NoStop}%
\bibitem [{\citenamefont {Narevicius}\ and\ \citenamefont
  {{Raizen}}(2012)}]{narevicius2012}%
  \BibitemOpen
  \bibfield  {author} {\bibinfo {author} {\bibfnamefont {E.}~\bibnamefont
  {Narevicius}}\ and\ \bibinfo {author} {\bibfnamefont {M.~G.}\ \bibnamefont
  {{Raizen}}},\ }\href@noop {} {\bibfield  {journal} {\bibinfo  {journal}
  {Chem. {Rev.}}\ }\textbf {\bibinfo {volume} {112}},\ \bibinfo {pages} {4879}
  (\bibinfo {year} {2012})}\BibitemShut {NoStop}%
\bibitem [{\citenamefont {Hogan}\ \emph {et~al.}(2011)\citenamefont {Hogan},
  \citenamefont {Motsch},\ and\ \citenamefont {Merkt}}]{hogan11a}%
  \BibitemOpen
  \bibfield  {author} {\bibinfo {author} {\bibfnamefont {S.~D.}\ \bibnamefont
  {Hogan}}, \bibinfo {author} {\bibfnamefont {M.}~\bibnamefont {Motsch}}, \
  and\ \bibinfo {author} {\bibfnamefont {F.}~\bibnamefont {Merkt}},\ }\href
  {\doibase 10.1039/C1CP21733J} {\bibfield  {journal} {\bibinfo  {journal}
  {Phys. Chem. Chem. Phys.}\ }\textbf {\bibinfo {volume} {13}},\ \bibinfo
  {pages} {18705} (\bibinfo {year} {2011})}\BibitemShut {NoStop}%
\bibitem [{\citenamefont {Kandula}\ \emph {et~al.}(2010)\citenamefont
  {Kandula}, \citenamefont {{Gohle}}, \citenamefont {{Pinkert}}, \citenamefont
  {{Ubachs}},\ and\ \citenamefont {{Eikema}}}]{kandula2010}%
  \BibitemOpen
  \bibfield  {author} {\bibinfo {author} {\bibfnamefont {D.~Z.}\ \bibnamefont
  {Kandula}}, \bibinfo {author} {\bibfnamefont {C.}~\bibnamefont {{Gohle}}},
  \bibinfo {author} {\bibfnamefont {T.~J.}\ \bibnamefont {{Pinkert}}}, \bibinfo
  {author} {\bibfnamefont {W.}~\bibnamefont {{Ubachs}}}, \ and\ \bibinfo
  {author} {\bibfnamefont {K.~S.~E.}\ \bibnamefont {{Eikema}}},\ }\href@noop {}
  {\bibfield  {journal} {\bibinfo  {journal} {Phys. {Rev.} {Lett.}}\ }\textbf
  {\bibinfo {volume} {105}},\ \bibinfo {pages} {063001} (\bibinfo {year}
  {2010})}\BibitemShut {NoStop}%
\bibitem [{\citenamefont {Cing\"oz}\ \emph {et~al.}(2012)\citenamefont
  {Cing\"oz}, \citenamefont {{Yost}}, \citenamefont {{Allison}}, \citenamefont
  {{Ruehl}}, \citenamefont {{Fermann}}, \citenamefont {{Hartl}},\ and\
  \citenamefont {{Ye}}}]{cingoz2012}%
  \BibitemOpen
  \bibfield  {author} {\bibinfo {author} {\bibfnamefont {A.}~\bibnamefont
  {Cing\"oz}}, \bibinfo {author} {\bibfnamefont {D.~C.}\ \bibnamefont
  {{Yost}}}, \bibinfo {author} {\bibfnamefont {T.~K.}\ \bibnamefont
  {{Allison}}}, \bibinfo {author} {\bibfnamefont {A.}~\bibnamefont {{Ruehl}}},
  \bibinfo {author} {\bibfnamefont {M.~E.}\ \bibnamefont {{Fermann}}}, \bibinfo
  {author} {\bibfnamefont {I.}~\bibnamefont {{Hartl}}}, \ and\ \bibinfo
  {author} {\bibfnamefont {J.}~\bibnamefont {{Ye}}},\ }\href@noop {} {\bibfield
   {journal} {\bibinfo  {journal} {Nature}\ }\textbf {\bibinfo {volume}
  {482}},\ \bibinfo {pages} {68} (\bibinfo {year} {2012})}\BibitemShut
  {NoStop}%
\bibitem [{\citenamefont {Rydberg}(1890)}]{rydberg1890}%
  \BibitemOpen
  \bibfield  {author} {\bibinfo {author} {\bibfnamefont {J.~R.}\ \bibnamefont
  {Rydberg}},\ }\href@noop {} {\bibfield  {journal} {\bibinfo  {journal}
  {Philos. {Mag.} {Series} 5}\ }\textbf {\bibinfo {volume} {29}},\ \bibinfo
  {pages} {331} (\bibinfo {year} {1890})}\BibitemShut {NoStop}%
\bibitem [{\citenamefont {Ritz}(1908)}]{ritz1908}%
  \BibitemOpen
  \bibfield  {author} {\bibinfo {author} {\bibfnamefont {W.}~\bibnamefont
  {Ritz}},\ }\href {\doibase 10.1086/141591} {\bibfield  {journal} {\bibinfo
  {journal} {\apj}\ }\textbf {\bibinfo {volume} {28}},\ \bibinfo {pages} {237}
  (\bibinfo {year} {1908})}\BibitemShut {NoStop}%
\bibitem [{\citenamefont {Mack}\ \emph {et~al.}(2011)\citenamefont {Mack},
  \citenamefont {{Karlewski}}, \citenamefont {{Hattermann}}, \citenamefont
  {{H}{\"o}ckh}, \citenamefont {{Jessen}}, \citenamefont {{Cano}},\ and\
  \citenamefont {{Fort}{\'a}gh}}]{mack2011}%
  \BibitemOpen
  \bibfield  {author} {\bibinfo {author} {\bibfnamefont {M.}~\bibnamefont
  {Mack}}, \bibinfo {author} {\bibfnamefont {F.}~\bibnamefont {{Karlewski}}},
  \bibinfo {author} {\bibfnamefont {H.}~\bibnamefont {{Hattermann}}}, \bibinfo
  {author} {\bibfnamefont {S.}~\bibnamefont {{H}{\"o}ckh}}, \bibinfo {author}
  {\bibfnamefont {F.}~\bibnamefont {{Jessen}}}, \bibinfo {author}
  {\bibfnamefont {D.}~\bibnamefont {{Cano}}}, \ and\ \bibinfo {author}
  {\bibfnamefont {J.}~\bibnamefont {{Fort}{\'a}gh}},\ }\href@noop {} {\bibfield
   {journal} {\bibinfo  {journal} {Phys. {Rev.} {A}}\ }\textbf {\bibinfo
  {volume} {83}},\ \bibinfo {pages} {052515} (\bibinfo {year}
  {2011})}\BibitemShut {NoStop}%
\bibitem [{\citenamefont {Udem}\ \emph {et~al.}(1999)\citenamefont {Udem},
  \citenamefont {{Reichert}}, \citenamefont {{Holzwarth}},\ and\ \citenamefont
  {{H}{\"a}nsch}}]{udem1999}%
  \BibitemOpen
  \bibfield  {author} {\bibinfo {author} {\bibfnamefont {T.}~\bibnamefont
  {Udem}}, \bibinfo {author} {\bibfnamefont {J.}~\bibnamefont {{Reichert}}},
  \bibinfo {author} {\bibfnamefont {R.}~\bibnamefont {{Holzwarth}}}, \ and\
  \bibinfo {author} {\bibfnamefont {T.~W.}\ \bibnamefont {{H}{\"a}nsch}},\
  }\href@noop {} {\bibfield  {journal} {\bibinfo  {journal} {Phys. {Rev.}
  {Lett.}}\ }\textbf {\bibinfo {volume} {82}},\ \bibinfo {pages} {3568}
  (\bibinfo {year} {1999})}\BibitemShut {NoStop}%
\bibitem [{\citenamefont {Sansonetti}\ \emph {et~al.}(1981)\citenamefont
  {Sansonetti}, \citenamefont {{Andrew}},\ and\ \citenamefont
  {{Verges}}}]{sansonetti1981}%
  \BibitemOpen
  \bibfield  {author} {\bibinfo {author} {\bibfnamefont {C.~J.}\ \bibnamefont
  {Sansonetti}}, \bibinfo {author} {\bibfnamefont {K.~L.}\ \bibnamefont
  {{Andrew}}}, \ and\ \bibinfo {author} {\bibfnamefont {J.}~\bibnamefont
  {{Verges}}},\ }\href@noop {} {\bibfield  {journal} {\bibinfo  {journal} {J.
  {Opt.} {Soc.} {Am.}}\ }\textbf {\bibinfo {volume} {71}},\ \bibinfo {pages}
  {423} (\bibinfo {year} {1981})}\BibitemShut {NoStop}%
\bibitem [{\citenamefont {Lorenzen}\ and\ \citenamefont
  {{Niemax}}(1984)}]{lorenzen1984}%
  \BibitemOpen
  \bibfield  {author} {\bibinfo {author} {\bibfnamefont {C.-J.}\ \bibnamefont
  {Lorenzen}}\ and\ \bibinfo {author} {\bibfnamefont {K.}~\bibnamefont
  {{Niemax}}},\ }\href@noop {} {\bibfield  {journal} {\bibinfo  {journal} {Z.
  {Phys.} {A}}\ }\textbf {\bibinfo {volume} {315}},\ \bibinfo {pages} {127}
  (\bibinfo {year} {1984})}\BibitemShut {NoStop}%
\bibitem [{\citenamefont {{O}'{Sullivan}}\ and\ \citenamefont
  {{Stoicheff}}(1983)}]{osullivan1983}%
  \BibitemOpen
  \bibfield  {author} {\bibinfo {author} {\bibfnamefont {M.~S.}\ \bibnamefont
  {{O}'{Sullivan}}}\ and\ \bibinfo {author} {\bibfnamefont {B.~P.}\
  \bibnamefont {{Stoicheff}}},\ }\href@noop {} {\bibfield  {journal} {\bibinfo
  {journal} {Can. {J.} {Phys.}}\ }\textbf {\bibinfo {volume} {61}},\ \bibinfo
  {pages} {940} (\bibinfo {year} {1983})}\BibitemShut {NoStop}%
\bibitem [{\citenamefont {Goy}\ \emph {et~al.}(1982)\citenamefont {Goy},
  \citenamefont {Raimond}, \citenamefont {Vitrant},\ and\ \citenamefont
  {Haroche}}]{goy1982}%
  \BibitemOpen
  \bibfield  {author} {\bibinfo {author} {\bibfnamefont {P.}~\bibnamefont
  {Goy}}, \bibinfo {author} {\bibfnamefont {J.~M.}\ \bibnamefont {Raimond}},
  \bibinfo {author} {\bibfnamefont {G.}~\bibnamefont {Vitrant}}, \ and\
  \bibinfo {author} {\bibfnamefont {S.}~\bibnamefont {Haroche}},\ }\href
  {\doibase 10.1103/PhysRevA.26.2733} {\bibfield  {journal} {\bibinfo
  {journal} {Phys. Rev. A}\ }\textbf {\bibinfo {volume} {26}},\ \bibinfo
  {pages} {2733} (\bibinfo {year} {1982})}\BibitemShut {NoStop}%
\bibitem [{\citenamefont {Sa{\ss}mannshausen}\ \emph
  {et~al.}(2013)\citenamefont {Sa{\ss}mannshausen}, \citenamefont {Merkt},\
  and\ \citenamefont {Deiglmayr}}]{sassmannshausen2013}%
  \BibitemOpen
  \bibfield  {author} {\bibinfo {author} {\bibfnamefont {H.}~\bibnamefont
  {Sa{\ss}mannshausen}}, \bibinfo {author} {\bibfnamefont {F.}~\bibnamefont
  {Merkt}}, \ and\ \bibinfo {author} {\bibfnamefont {J.}~\bibnamefont
  {Deiglmayr}},\ }\href {\doibase 10.1103/PhysRevA.87.032519} {\bibfield
  {journal} {\bibinfo  {journal} {Phys. Rev. A}\ }\textbf {\bibinfo {volume}
  {87}},\ \bibinfo {pages} {032519} (\bibinfo {year} {2013})}\BibitemShut
  {NoStop}%
\bibitem [{\citenamefont {Deiglmayr}\ \emph {et~al.}(2014)\citenamefont
  {Deiglmayr}, \citenamefont {Sa{\ss}mannshausen}, \citenamefont {Pillet},\
  and\ \citenamefont {Merkt}}]{deiglmayr2014}%
  \BibitemOpen
  \bibfield  {author} {\bibinfo {author} {\bibfnamefont {J.}~\bibnamefont
  {Deiglmayr}}, \bibinfo {author} {\bibfnamefont {H.}~\bibnamefont
  {Sa{\ss}mannshausen}}, \bibinfo {author} {\bibfnamefont {P.}~\bibnamefont
  {Pillet}}, \ and\ \bibinfo {author} {\bibfnamefont {F.}~\bibnamefont
  {Merkt}},\ }\href@noop {} {\bibfield  {journal} {\bibinfo  {journal} {Phys.
  Rev. Lett.}\ }\textbf {\bibinfo {volume} {113}},\ \bibinfo {pages} {193001}
  (\bibinfo {year} {2014})}\BibitemShut {NoStop}%
\bibitem [{\citenamefont {Sa{\ss}mannshausen}\ \emph
  {et~al.}(2015)\citenamefont {Sa{\ss}mannshausen}, \citenamefont {Merkt},\
  and\ \citenamefont {Deiglmayr}}]{sassmannshausen2015}%
  \BibitemOpen
  \bibfield  {author} {\bibinfo {author} {\bibfnamefont {H.}~\bibnamefont
  {Sa{\ss}mannshausen}}, \bibinfo {author} {\bibfnamefont {F.}~\bibnamefont
  {Merkt}}, \ and\ \bibinfo {author} {\bibfnamefont {J.}~\bibnamefont
  {Deiglmayr}},\ }\href {\doibase 10.1103/PhysRevLett.114.133201} {\bibfield
  {journal} {\bibinfo  {journal} {Phys. Rev. Lett.}\ }\textbf {\bibinfo
  {volume} {114}},\ \bibinfo {pages} {133201} (\bibinfo {year}
  {2015})}\BibitemShut {NoStop}%
\bibitem [{\citenamefont {Raimond}\ \emph {et~al.}(1978)\citenamefont
  {Raimond}, \citenamefont {Gross}, \citenamefont {Fabre}, \citenamefont
  {Haroche},\ and\ \citenamefont {Stroke}}]{raimond1978}%
  \BibitemOpen
  \bibfield  {author} {\bibinfo {author} {\bibfnamefont {J.~M.}\ \bibnamefont
  {Raimond}}, \bibinfo {author} {\bibfnamefont {M.}~\bibnamefont {Gross}},
  \bibinfo {author} {\bibfnamefont {C.}~\bibnamefont {Fabre}}, \bibinfo
  {author} {\bibfnamefont {S.}~\bibnamefont {Haroche}}, \ and\ \bibinfo
  {author} {\bibfnamefont {H.~H.}\ \bibnamefont {Stroke}},\ }\href {\doibase
  10.1088/0022-3700/11/24/004} {\bibfield  {journal} {\bibinfo  {journal} {J.
  Phys. B}\ }\textbf {\bibinfo {volume} {11}},\ \bibinfo {pages} {L765}
  (\bibinfo {year} {1978})}\BibitemShut {NoStop}%
\bibitem [{\citenamefont {Beterov}\ \emph {et~al.}(2007)\citenamefont
  {Beterov}, \citenamefont {Tretyakov}, \citenamefont {Ryabtsev}, \citenamefont
  {Ekers},\ and\ \citenamefont {Bezuglov}}]{beterov2007}%
  \BibitemOpen
  \bibfield  {author} {\bibinfo {author} {\bibfnamefont {I.~I.}\ \bibnamefont
  {Beterov}}, \bibinfo {author} {\bibfnamefont {D.~B.}\ \bibnamefont
  {Tretyakov}}, \bibinfo {author} {\bibfnamefont {I.~I.}\ \bibnamefont
  {Ryabtsev}}, \bibinfo {author} {\bibfnamefont {A.}~\bibnamefont {Ekers}}, \
  and\ \bibinfo {author} {\bibfnamefont {N.~N.}\ \bibnamefont {Bezuglov}},\
  }\href@noop {} {\bibfield  {journal} {\bibinfo  {journal} {Phys. Rev. A}\
  }\textbf {\bibinfo {volume} {75}},\ \bibinfo {pages} {052720} (\bibinfo
  {year} {2007})}\BibitemShut {NoStop}%
\bibitem [{\citenamefont {Drever}\ \emph {et~al.}(1983)\citenamefont {Drever},
  \citenamefont {Hall}, \citenamefont {Kowalski}, \citenamefont {Hough},
  \citenamefont {Ford}, \citenamefont {Munley},\ and\ \citenamefont
  {Ward}}]{drever1983}%
  \BibitemOpen
  \bibfield  {author} {\bibinfo {author} {\bibfnamefont {R.~W.~P.}\
  \bibnamefont {Drever}}, \bibinfo {author} {\bibfnamefont {J.~L.}\
  \bibnamefont {Hall}}, \bibinfo {author} {\bibfnamefont {F.~V.}\ \bibnamefont
  {Kowalski}}, \bibinfo {author} {\bibfnamefont {J.}~\bibnamefont {Hough}},
  \bibinfo {author} {\bibfnamefont {G.~M.}\ \bibnamefont {Ford}}, \bibinfo
  {author} {\bibfnamefont {A.~J.}\ \bibnamefont {Munley}}, \ and\ \bibinfo
  {author} {\bibfnamefont {H.}~\bibnamefont {Ward}},\ }\href@noop {} {\bibfield
   {journal} {\bibinfo  {journal} {Appl. Phys. B}\ }\textbf {\bibinfo {volume}
  {31}},\ \bibinfo {pages} {97} (\bibinfo {year} {1983})}\BibitemShut {NoStop}%
\bibitem [{\citenamefont {Haubrich}\ and\ \citenamefont
  {Wynands}(1996)}]{haubrich1996}%
  \BibitemOpen
  \bibfield  {author} {\bibinfo {author} {\bibfnamefont {D.}~\bibnamefont
  {Haubrich}}\ and\ \bibinfo {author} {\bibfnamefont {R.}~\bibnamefont
  {Wynands}},\ }\href@noop {} {\bibfield  {journal} {\bibinfo  {journal} {Opt.
  Comm.}\ }\textbf {\bibinfo {volume} {123}},\ \bibinfo {pages} {558} (\bibinfo
  {year} {1996})}\BibitemShut {NoStop}%
\bibitem [{\citenamefont {Sanguinetti}\ \emph {et~al.}(2009)\citenamefont
  {Sanguinetti}, \citenamefont {Majeed}, \citenamefont {Jones},\ and\
  \citenamefont {Varcoe}}]{sanguinetti2009}%
  \BibitemOpen
  \bibfield  {author} {\bibinfo {author} {\bibfnamefont {B.}~\bibnamefont
  {Sanguinetti}}, \bibinfo {author} {\bibfnamefont {H.~O.}\ \bibnamefont
  {Majeed}}, \bibinfo {author} {\bibfnamefont {M.~L.}\ \bibnamefont {Jones}}, \
  and\ \bibinfo {author} {\bibfnamefont {B.~T.~H.}\ \bibnamefont {Varcoe}},\
  }\href@noop {} {\bibfield  {journal} {\bibinfo  {journal} {J. Phys. B}\
  }\textbf {\bibinfo {volume} {42}},\ \bibinfo {pages} {165004} (\bibinfo
  {year} {2009})}\BibitemShut {NoStop}%
\bibitem [{\citenamefont {Kubina}\ \emph {et~al.}(2005)\citenamefont {Kubina},
  \citenamefont {{Adel}}, \citenamefont {{Adler}}, \citenamefont {{Grosche}},
  \citenamefont {H\"ansch}, \citenamefont {{Holzwarth}}, \citenamefont
  {{Leitenstorfer}}, \citenamefont {{Lipphardt}},\ and\ \citenamefont
  {{Schnatz}}}]{kubina2005}%
  \BibitemOpen
  \bibfield  {author} {\bibinfo {author} {\bibfnamefont {P.}~\bibnamefont
  {Kubina}}, \bibinfo {author} {\bibfnamefont {P.}~\bibnamefont {{Adel}}},
  \bibinfo {author} {\bibfnamefont {F.}~\bibnamefont {{Adler}}}, \bibinfo
  {author} {\bibfnamefont {G.}~\bibnamefont {{Grosche}}}, \bibinfo {author}
  {\bibfnamefont {T.~W.}\ \bibnamefont {H\"ansch}}, \bibinfo {author}
  {\bibfnamefont {R.}~\bibnamefont {{Holzwarth}}}, \bibinfo {author}
  {\bibfnamefont {A.}~\bibnamefont {{Leitenstorfer}}}, \bibinfo {author}
  {\bibfnamefont {B.}~\bibnamefont {{Lipphardt}}}, \ and\ \bibinfo {author}
  {\bibfnamefont {H.}~\bibnamefont {{Schnatz}}},\ }\href@noop {} {\bibfield
  {journal} {\bibinfo  {journal} {Opt. Expr.}\ }\textbf {\bibinfo {volume}
  {13}},\ \bibinfo {pages} {904} (\bibinfo {year} {2005})}\BibitemShut
  {NoStop}%
\bibitem [{\citenamefont {Ovsiannikov}\ \emph {et~al.}(2011)\citenamefont
  {Ovsiannikov}, \citenamefont {Glukhov},\ and\ \citenamefont
  {Nekipelov}}]{ovsiannikov2011}%
  \BibitemOpen
  \bibfield  {author} {\bibinfo {author} {\bibfnamefont {V.~D.}\ \bibnamefont
  {Ovsiannikov}}, \bibinfo {author} {\bibfnamefont {I.~L.}\ \bibnamefont
  {Glukhov}}, \ and\ \bibinfo {author} {\bibfnamefont {E.~A.}\ \bibnamefont
  {Nekipelov}},\ }\href {\doibase 10.1088/0953-4075/44/19/195010} {\bibfield
  {journal} {\bibinfo  {journal} {J. Phys. B}\ }\textbf {\bibinfo {volume}
  {44}},\ \bibinfo {pages} {195010} (\bibinfo {year} {2011})}\BibitemShut
  {NoStop}%
\bibitem [{\citenamefont {Gallagher}(2005)}]{gallagher2005}%
  \BibitemOpen
  \bibfield  {author} {\bibinfo {author} {\bibfnamefont {T.~F.}\ \bibnamefont
  {Gallagher}},\ }\href@noop {} {\emph {\bibinfo {title} {Rydberg Atoms}}}\
  (\bibinfo  {publisher} {Cambridge University Press},\ \bibinfo {year}
  {2005})\BibitemShut {NoStop}%
\bibitem [{\citenamefont {Arimondo}\ \emph {et~al.}(1977)\citenamefont
  {Arimondo}, \citenamefont {Inguscio},\ and\ \citenamefont
  {Violino}}]{Arimondo1977}%
  \BibitemOpen
  \bibfield  {author} {\bibinfo {author} {\bibfnamefont {E.}~\bibnamefont
  {Arimondo}}, \bibinfo {author} {\bibfnamefont {M.}~\bibnamefont {Inguscio}},
  \ and\ \bibinfo {author} {\bibfnamefont {P.}~\bibnamefont {Violino}},\
  }\href@noop {} {\bibfield  {journal} {\bibinfo  {journal} {Rev. Mod. Phys.}\
  }\textbf {\bibinfo {volume} {49}},\ \bibinfo {pages} {31} (\bibinfo {year}
  {1977})}\BibitemShut {NoStop}%
\bibitem [{\citenamefont {Felinger}(1998)}]{felinger1998}%
  \BibitemOpen
  \bibfield  {author} {\bibinfo {author} {\bibfnamefont {A.}~\bibnamefont
  {Felinger}},\ }\href@noop {} {\emph {\bibinfo {title} {{Data Analysis and
  Signal Processing in Chromatography/ Data Handling in Science and
  Technology}}}},\ Vol.~\bibinfo {volume} {21}\ (\bibinfo  {publisher}
  {Elsevier, Amsterdam, the Netherlands},\ \bibinfo {year} {1998})\ p.\
  \bibinfo {pages} {69ff}\BibitemShut {NoStop}%
\bibitem [{\citenamefont {Liberman}\ \emph {et~al.}(1983)\citenamefont
  {Liberman}, \citenamefont {Pinard},\ and\ \citenamefont
  {Taleb}}]{liberman1983}%
  \BibitemOpen
  \bibfield  {author} {\bibinfo {author} {\bibfnamefont {S.}~\bibnamefont
  {Liberman}}, \bibinfo {author} {\bibfnamefont {J.}~\bibnamefont {Pinard}}, \
  and\ \bibinfo {author} {\bibfnamefont {A.}~\bibnamefont {Taleb}},\
  }\href@noop {} {\bibfield  {journal} {\bibinfo  {journal} {Phys. Rev. Lett.}\
  }\textbf {\bibinfo {volume} {50}},\ \bibinfo {pages} {888} (\bibinfo {year}
  {1983})}\BibitemShut {NoStop}%
\bibitem [{\citenamefont {{Grimm}}\ \emph {et~al.}(2000)\citenamefont
  {{Grimm}}, \citenamefont {{Weidem{\"u}ller}},\ and\ \citenamefont
  {{Ovchinnikov}}}]{grimm2000}%
  \BibitemOpen
  \bibfield  {author} {\bibinfo {author} {\bibfnamefont {R.}~\bibnamefont
  {{Grimm}}}, \bibinfo {author} {\bibfnamefont {M.}~\bibnamefont
  {{Weidem{\"u}ller}}}, \ and\ \bibinfo {author} {\bibfnamefont {Y.~B.}\
  \bibnamefont {{Ovchinnikov}}},\ }in\ \href@noop {} {\emph {\bibinfo
  {booktitle} {{Adv. At. Mol. Opt. Phys.}}}},\ Vol.~\bibinfo {volume} {42},\
  \bibinfo {editor} {edited by\ \bibinfo {editor} {\bibfnamefont {B.~B.}\
  \bibnamefont {{Walther}}}\ and\ \bibinfo {editor} {\bibnamefont {Herbert}}}\
  (\bibinfo  {publisher} {{Academic Press}},\ \bibinfo {year} {2000})\
  p.~\bibinfo {pages} {95}\BibitemShut {NoStop}%
\bibitem [{\citenamefont {Hollberg}\ and\ \citenamefont
  {Hall}(1984)}]{hollberg1984}%
  \BibitemOpen
  \bibfield  {author} {\bibinfo {author} {\bibfnamefont {L.}~\bibnamefont
  {Hollberg}}\ and\ \bibinfo {author} {\bibfnamefont {J.~L.}\ \bibnamefont
  {Hall}},\ }\href@noop {} {\bibfield  {journal} {\bibinfo  {journal} {Phys.
  Rev. Lett.}\ }\textbf {\bibinfo {volume} {53}},\ \bibinfo {pages} {230}
  (\bibinfo {year} {1984})}\BibitemShut {NoStop}%
\bibitem [{\citenamefont {Amaldi}\ and\ \citenamefont
  {Segr\`e}(1934)}]{Amaldi1934}%
  \BibitemOpen
  \bibfield  {author} {\bibinfo {author} {\bibfnamefont {E.}~\bibnamefont
  {Amaldi}}\ and\ \bibinfo {author} {\bibfnamefont {E.}~\bibnamefont
  {Segr\`e}},\ }\href@noop {} {\bibfield  {journal} {\bibinfo  {journal} {Il
  Nuovo Cimento}\ }\textbf {\bibinfo {volume} {11}},\ \bibinfo {pages} {145}
  (\bibinfo {year} {1934})}\BibitemShut {NoStop}%
\bibitem [{\citenamefont {Fermi}(1934)}]{Fermi1934}%
  \BibitemOpen
  \bibfield  {author} {\bibinfo {author} {\bibfnamefont {E.}~\bibnamefont
  {Fermi}},\ }\href {\doibase 10.1007/BF02959829} {\bibfield  {journal}
  {\bibinfo  {journal} {Il Nuovo Cimento}\ }\textbf {\bibinfo {volume} {11}},\
  \bibinfo {pages} {157} (\bibinfo {year} {1934})}\BibitemShut {NoStop}%
\bibitem [{\citenamefont {Bahrim}\ \emph {et~al.}(2001)\citenamefont {Bahrim},
  \citenamefont {Thumm},\ and\ \citenamefont {Fabrikant}}]{bahrim2001}%
  \BibitemOpen
  \bibfield  {author} {\bibinfo {author} {\bibfnamefont {C.}~\bibnamefont
  {Bahrim}}, \bibinfo {author} {\bibfnamefont {U.}~\bibnamefont {Thumm}}, \
  and\ \bibinfo {author} {\bibfnamefont {I.~I.}\ \bibnamefont {Fabrikant}},\
  }\href@noop {} {\bibfield  {journal} {\bibinfo  {journal} {J. Phys. B}\
  }\textbf {\bibinfo {volume} {34}},\ \bibinfo {pages} {L195} (\bibinfo {year}
  {2001})}\BibitemShut {NoStop}%
\bibitem [{\citenamefont {Gallagher}\ and\ \citenamefont
  {Pillet}(2008)}]{gallagher2008}%
  \BibitemOpen
  \bibfield  {author} {\bibinfo {author} {\bibfnamefont {T.~F.}\ \bibnamefont
  {Gallagher}}\ and\ \bibinfo {author} {\bibfnamefont {P.}~\bibnamefont
  {Pillet}},\ }in\ \href@noop {} {\emph {\bibinfo {booktitle} {Adv. At. Mol.
  Opt. Phys.}}},\ Vol.~\bibinfo {volume} {56},\ \bibinfo {editor} {edited by\
  \bibinfo {editor} {\bibfnamefont {E.}~\bibnamefont {Arimondo}}}\ (\bibinfo
  {publisher} {Academic Press},\ \bibinfo {year} {2008})\ pp.\ \bibinfo {pages}
  {161--218}\BibitemShut {NoStop}%
\bibitem [{\citenamefont {{P. J. Mohr}}\ \emph {et~al.}(2015)\citenamefont {{P.
  J. Mohr}}, \citenamefont {{B. N. Taylor}},\ and\ \citenamefont {{D. B.
  Newell}}}]{Nist2014}%
  \BibitemOpen
  \bibfield  {author} {\bibinfo {author} {\bibnamefont {{P. J. Mohr}}},
  \bibinfo {author} {\bibnamefont {{B. N. Taylor}}}, \ and\ \bibinfo {author}
  {\bibnamefont {{D. B. Newell}}},\ }\href@noop {} {\enquote {\bibinfo {title}
  {{The 2014 CODATA Recommended Values of the Fundamental Physical Constants
  (Web Version 7.0). This database was developed by J. Baker, M. Douma, and S.
  Kotochigova. National Institute of Standards and Technology, Gaithersburg, MD
  20899}},}\ }\bibinfo {howpublished} {\url{http://physics.nist.gov/constants}}
  (\bibinfo {year} {2015})\BibitemShut {NoStop}%
\bibitem [{\citenamefont {{CIAAW}}(2015)}]{CIAAW2015}%
  \BibitemOpen
  \bibfield  {author} {\bibinfo {author} {\bibnamefont {{CIAAW}}},\ }\href@noop
  {} {\enquote {\bibinfo {title} {{Atomic weights of the elements 2015}},}\
  }\bibinfo {howpublished} {\url{http://ciaaw.org/atomic-weights}} (\bibinfo
  {year} {2015})\BibitemShut {NoStop}%
\bibitem [{\citenamefont {Drake}\ and\ \citenamefont
  {Swainson}(1991)}]{drake_1991}%
  \BibitemOpen
  \bibfield  {author} {\bibinfo {author} {\bibfnamefont {G.~W.~F.}\
  \bibnamefont {Drake}}\ and\ \bibinfo {author} {\bibfnamefont {R.~A.}\
  \bibnamefont {Swainson}},\ }\href@noop {} {\bibfield  {journal} {\bibinfo
  {journal} {Phys. Rev. A}\ }\textbf {\bibinfo {volume} {44}},\ \bibinfo
  {pages} {5448} (\bibinfo {year} {1991})}\BibitemShut {NoStop}%
\bibitem [{\citenamefont {Seiler}\ \emph {et~al.}(2011)\citenamefont {Seiler},
  \citenamefont {{Hogan}}, \citenamefont {{Schmutz}}, \citenamefont {{Agner}},\
  and\ \citenamefont {{Merkt}}}]{seiler2011b}%
  \BibitemOpen
  \bibfield  {author} {\bibinfo {author} {\bibfnamefont {C.}~\bibnamefont
  {Seiler}}, \bibinfo {author} {\bibfnamefont {S.~D.}\ \bibnamefont {{Hogan}}},
  \bibinfo {author} {\bibfnamefont {H.}~\bibnamefont {{Schmutz}}}, \bibinfo
  {author} {\bibfnamefont {J.~A.}\ \bibnamefont {{Agner}}}, \ and\ \bibinfo
  {author} {\bibfnamefont {F.}~\bibnamefont {{Merkt}}},\ }\href@noop {}
  {\bibfield  {journal} {\bibinfo  {journal} {Phys. {Rev.} {Lett.}}\ }\textbf
  {\bibinfo {volume} {106}},\ \bibinfo {pages} {073003} (\bibinfo {year}
  {2011})}\BibitemShut {NoStop}%
\bibitem [{\citenamefont {Motsch}\ \emph {et~al.}(2014)\citenamefont {Motsch},
  \citenamefont {{Jansen}}, \citenamefont {{Agner}}, \citenamefont
  {{Schmutz}},\ and\ \citenamefont {{Merkt}}}]{motsch2014}%
  \BibitemOpen
  \bibfield  {author} {\bibinfo {author} {\bibfnamefont {M.}~\bibnamefont
  {Motsch}}, \bibinfo {author} {\bibfnamefont {P.}~\bibnamefont {{Jansen}}},
  \bibinfo {author} {\bibfnamefont {J.~A.}\ \bibnamefont {{Agner}}}, \bibinfo
  {author} {\bibfnamefont {H.}~\bibnamefont {{Schmutz}}}, \ and\ \bibinfo
  {author} {\bibfnamefont {F.}~\bibnamefont {{Merkt}}},\ }\href@noop {}
  {\bibfield  {journal} {\bibinfo  {journal} {Phys. {Rev.} {A}}\ }\textbf
  {\bibinfo {volume} {89}},\ \bibinfo {pages} {043420} (\bibinfo {year}
  {2014})}\BibitemShut {NoStop}%
\bibitem [{\citenamefont {Liu}\ \emph {et~al.}(2009)\citenamefont {Liu},
  \citenamefont {{Salumbides}}, \citenamefont {{Hollenstein}}, \citenamefont
  {{Koelemeij}}, \citenamefont {{Eikema}}, \citenamefont {{Ubachs}},\ and\
  \citenamefont {{Merkt}}}]{liu2009}%
  \BibitemOpen
  \bibfield  {author} {\bibinfo {author} {\bibfnamefont {J.}~\bibnamefont
  {Liu}}, \bibinfo {author} {\bibfnamefont {E.~J.}\ \bibnamefont
  {{Salumbides}}}, \bibinfo {author} {\bibfnamefont {U.}~\bibnamefont
  {{Hollenstein}}}, \bibinfo {author} {\bibfnamefont {J.~C.~J.}\ \bibnamefont
  {{Koelemeij}}}, \bibinfo {author} {\bibfnamefont {K.~S.~E.}\ \bibnamefont
  {{Eikema}}}, \bibinfo {author} {\bibfnamefont {W.}~\bibnamefont {{Ubachs}}},
  \ and\ \bibinfo {author} {\bibfnamefont {F.}~\bibnamefont {{Merkt}}},\
  }\href@noop {} {\bibfield  {journal} {\bibinfo  {journal} {{J.} {Chem.}
  {Phys.}}\ }\textbf {\bibinfo {volume} {130}},\ \bibinfo {pages} {174306}
  (\bibinfo {year} {2009})}\BibitemShut {NoStop}%
\bibitem [{\citenamefont {Piszczatowski}\ \emph {et~al.}(2009)\citenamefont
  {Piszczatowski}, \citenamefont {{\L}ach}, \citenamefont {{Przybytek}},
  \citenamefont {{Komasa}}, \citenamefont {{Pachucki}},\ and\ \citenamefont
  {{Jeziorski}}}]{piszczatowski2009}%
  \BibitemOpen
  \bibfield  {author} {\bibinfo {author} {\bibfnamefont {K.}~\bibnamefont
  {Piszczatowski}}, \bibinfo {author} {\bibfnamefont {G.}~\bibnamefont
  {{\L}ach}}, \bibinfo {author} {\bibfnamefont {M.}~\bibnamefont
  {{Przybytek}}}, \bibinfo {author} {\bibfnamefont {J.}~\bibnamefont
  {{Komasa}}}, \bibinfo {author} {\bibfnamefont {K.}~\bibnamefont
  {{Pachucki}}}, \ and\ \bibinfo {author} {\bibfnamefont {B.}~\bibnamefont
  {{Jeziorski}}},\ }\href@noop {} {\bibfield  {journal} {\bibinfo  {journal}
  {J. {Chem.} {Theo.} {Comp.}}\ }\textbf {\bibinfo {volume} {5}},\ \bibinfo
  {pages} {3039} (\bibinfo {year} {2009})}\BibitemShut {NoStop}%
\bibitem [{\citenamefont {Pachucki}\ and\ \citenamefont
  {{Komasa}}(2010)}]{pachucki2010}%
  \BibitemOpen
  \bibfield  {author} {\bibinfo {author} {\bibfnamefont {K.}~\bibnamefont
  {Pachucki}}\ and\ \bibinfo {author} {\bibfnamefont {J.}~\bibnamefont
  {{Komasa}}},\ }\href@noop {} {\bibfield  {journal} {\bibinfo  {journal}
  {Phys. {Chem.} {Chem.} {Phys.}}\ }\textbf {\bibinfo {volume} {12}},\ \bibinfo
  {pages} {9188} (\bibinfo {year} {2010})}\BibitemShut {NoStop}%
\bibitem [{\citenamefont {Pachucki}\ and\ \citenamefont
  {{Komasa}}(2014)}]{pachucki2014}%
  \BibitemOpen
  \bibfield  {author} {\bibinfo {author} {\bibfnamefont {K.}~\bibnamefont
  {Pachucki}}\ and\ \bibinfo {author} {\bibfnamefont {J.}~\bibnamefont
  {{Komasa}}},\ }\href@noop {} {\bibfield  {journal} {\bibinfo  {journal} {{J.}
  {Chem.} {Phys.}}\ }\textbf {\bibinfo {volume} {141}},\ \bibinfo {pages}
  {224103} (\bibinfo {year} {2014})}\BibitemShut {NoStop}%
\end{thebibliography}

%

\end{document}